\documentclass[twocolumn]{aastex631}
\usepackage{CJK}
\usepackage{amsfonts}
\usepackage{amsmath}
\usepackage{mathrsfs}
\usepackage{epsfig}
\usepackage{subfigure}

\def\nuflub8{\phi^\nu_B}
\def\nuflube7{\phi^\nu_{Be}}

\def\kpc{\,\mathrm{kpc}}
\def\km{\,\mathrm{km}}
\def\m{\,\mathrm{m}}

\def\GeV{\,\mathrm{GeV}}

\def\GV{\,\mathrm{GV}}

\def\cm{\,\mathrm{cm}}
\def\s{\,\mathrm{s}}
\def\p{\,\mathrm{p}}

\def\C{\,\mathrm{C}}
\def\N{\,\mathrm{N}}
\def\O{\,\mathrm{O}}

\def\B{\,\mathrm{B}}
\def\A{\,\mathrm{A}}
\def\sr{\,\mathrm{sr}}

\def\Rbr{\,R_{\mathrm{br}}}
\def\kpc{\,\mathrm{kpc}}
\def\km{\,\mathrm{km}}
\def\m{\,\mathrm{m}}

\def\GeV{\,\mathrm{GeV}}

\def\GV{\,\mathrm{GV}}

\def\cm{\,\mathrm{cm}}
\def\s{\,\mathrm{s}}
\def\p{\,\mathrm{p}}

\def\C{\,\mathrm{C}}
\def\N{\,\mathrm{N}}
\def\O{\,\mathrm{O}}

\def\B{\,\mathrm{B}}
\def\A{\,\mathrm{A}}
\def\sr{\,\mathrm{sr}}

\def\Rbr{\,R_{\mathrm{br}}}

\def\R{\mathcal{R}}

\def\kpc{\,\mathrm{kpc}}
\def\km{\,\mathrm{km}}

\def\cm{\,\mathrm{cm}}
\def\s{\,\mathrm{s}}
\def\sr{\,\mathrm{sr}}
\newcolumntype{p}{D{,}{\pm}{-1}}

\def\c{\,\mathrm{c}}

\def\g{\,\mathrm{g}}

\shorttitle{Hybrid origins of the CR spectral hardening}
\shortauthors{Niu}

\begin{document}
\begin{CJK*}{UTF8}{gbsn}

\title{Hybrid origins of the cosmic-ray nuclei spectral hardening at a few hundred GV}

\correspondingauthor{Jia-Shu Niu}
\email{jsniu@sxu.edu.cn}

\author[0000-0001-5232-9500]{Jia-Shu Niu (牛家树)}
\affil{Institute of Theoretical Physics, Shanxi University, Taiyuan 030006, China}
\affil{State Key Laboratory of Quantum Optics and Quantum Optics Devices, Shanxi University, Taiyuan 030006, China}

\begin{abstract}
  Many experiments have confirmed the spectral hardening at a few hundred GV of cosmic-ray (CR) nuclei spectra, and three general different origins have been proposed: the primary source acceleration, the propagation, and the superposition of different kinds of sources. The AMS-02 CR nuclei spectra of He, C, N, O, Ne, Mg, Si, and B (which including B and its dominating parents species) are collected to study the necessity of employing a break in diffusion coefficient and independent breaks in primary source injection spectra to reproduce the spectral hardening at a few hundred GV. For comparison, three different schemes are introduced to do the global fitting. The fitting results show that both the break in diffusion coefficient and the independent breaks in primary source injection spectra are needed, which are corresponding to the spatial dependent propagation and the superposition of different kinds of sources, respectively. Consequently, the nuclei spectral hardening in a few hundred GV should have hybrid origins. Moreover, the CR spectral indices of He and Ne show large deviations to other species in low-rigidity region, which indicates their different CR origins. 
\end{abstract}

\section{Motivation}

Many space-borne and ground-based experiments have confirmed the spectral hardening at a few hundred GV in cosmic-ray (CR) nuclei species (such as ATIC-2 \citep{ATIC2006}, CREAM \citep{CREAM2010}, and PAMELA \citep{PAMELA2011}). 
The space station experiment Alpha Magnetic Spectrometer (AMS-02) improves the measurement precision of the CR fluxes by an order of magnitude of the systematics \citep{AMS2013} and leads us to a precision-driven era. The released spectra of different nuclei species by AMS-02 (including the primary CR species: proton \citep{AMS02_proton}, helium (He), carbon (C), oxygen (O) \citep{AMS02_He_C_O}, neon (Ne), magnesium (Mg), silicon (Si) \citep{AMS02_Ne_Mg_Si}, and iron (Fe) \citep{AMS02_Fe00}; the secondary CR species: lithium (Li), beryllium (Be), boron (B) \citep{AMS02_Li_Be_B}, and fluorine (F) \citep{AMS02_F00}; the hybrid CR species: nitrogen (N) \citep{AMS02_N}, sodium (Na), and aluminum (Al) \citep{AMS02_Na_Al00} provide us an excellent opportunity to study the origin, acceleration and propagation of CRs. As the most obvious and attractive fine structure in AMS-02 nuclei spectra, the spectral hardening in the region of $100-1000$ GV has been studied by many works.

One of the most promising scenarios (see, e.g.,  \citet{Blasi2012,Tomassetti2012,Tomassetti2015apjl01,Tomassetti2015prd,Feng2016,Genolini2017,Jin2016CPC,Guo2018cpc,Guo2018prd,Liu2018,Niu2019,Boschini2020apj,Boschini2020apjs}) is that the spectral hardening comes from the CR propagation process. Phenomenologically, in such scenario, the secondary nuclei spectra should harden even more than that of the primary ones at a few hundred GV\footnote{The secondary species spectra not only inherit the hardening from the primary species (which is caused by the propagation of primary species), but are also hardened by their own propagation processes.}, which is equivalent to add an extra high-rigidity break in the diffusion coefficient. Some previous works show that AMS-02 nuclei data favor the hardening coming from the propagation process rather than the CR primary source injections in a statistical meaning (see, e.g., \citet{Genolini2017,Niu2020}).

However, some recent works show that the propagation origin of the hardening can not be easily established. (see, e.g., \citet{Yuan2020jcap,Niu2021cpc}). Because the secondary CR species (such as Li, Be, and B) are produced in collisions of primary CR particles (such as C, N, and O) with interstellar medium (ISM), the spectral hardening of the secondary CR species inherits from that of the CR primary species. The test of such process should consider all the contributions from the parents species, at least the dominating ones.

In detail, the contribution of C to B flux is about 20\%, which is almost equal to N but less than O \citep{Genolini2018}. In \citet{Niu2021cpc}, it shows that not only the break rigidity (at a few hundred GV), but also the differences between the spectral index below and above the break of C, N, and O are different. In such a case, the conclusions obtained from B/C ratio alone cannot represent the real propagation process completely (such as in \citet{Genolini2017,Genolini2019}). 
Moreover, the spectra of proton and He have very small uncertainties because of the extremely large event number, if one uses these spectra in global fitting based on a uniform primary source injection for all the CR nuclei species, they dominate the injection spectra parameters and would seriously dilute the impacts of the real parents species (like that of  C, N, O, Ne, Mg, and Si) on the daughter species (like that of Li, Be, and B) (such as in \citet{Niu2019,Niu2020}). As a result, independent primary source injections are needed.

In this work, the AMS-02 CR nuclei spectra of C, N, O, Ne, Mg, and Si are used as the parents species, and that of B is used as the daughter species \footnote{The spectra of Li and Be are not used in this work because some recent works show that they might have extra primary components \citep{Boschini2020apj,Niu2019,Niu2020} and it needs to re-scale the production cross sections if we want to reproduce their spectra with that of B simutaneously \citep{Luque202103,Luque202107}.}. 
The spectrum of He is also included in the data set, which could provide us valuable comparisons with other species (especially C and O). 
This clean data set could not only help us to check the consistency between the observed data and the CR model, but also avoid the systematics between different experiments.

\section{Setups}

  In this work, we design three schemes to test the properties of the spectral hardening in the region of 100-1000 GV. In Scheme I, high-rigidity breaks are simultaneously employed in the diffusion coefficient (with one break) and primary source injection spectra for different species (with independent breaks); in Scheme II, independent high-rigidity breaks are employed in the primary source injection spectra for different species; in Scheme III, one high-rigidity break is employed in the diffusion coefficient in charge of the spectral hardening.

\subsection{Models for Different Schemes}

A modified version of the diffusion-reacceleration scenario is used to describe the propagation process \citep{Yuan2019SCPMA}, which could successfully reproduce the spectra in low-rigidity regions. For Scheme I and III, the diffusion coefficient includes a high-rigidity break and is parameterized as
\begin{equation}
  \label{eq:diffusion_coefficient_1}
  D_{xx}(R)=  D_{0} \cdot \beta^{\eta} \left(\frac{\Rbr}{R_{0}}\right) \times \left\{
    \begin{array}{ll}
      \left( \dfrac{R}{\Rbr} \right)^{\delta_{1}} & R \le \Rbr\\
      \left( \dfrac{R}{\Rbr} \right)^{\delta_{2}} & R > \Rbr
    \end{array}
  \right.,
\end{equation}
where $R\equiv pc/Ze$ is the rigidity, $\beta$ is the velocity of the particle in unit of light speed $c$, $\Rbr$ is the high-rigidity break, $\delta_{1}$ and $\delta_{2}$ are the diffusion slopes below and above the break, and $R_{0}$ is the reference rigidity (4 GV).
For Scheme II, the diffusion coefficient without the break is parameterized as
\begin{equation}
  \label{eq:diffusion_coefficient_2}
  D_{xx}(R)=  D_{0} \cdot \beta^{\eta} \left(\frac{\Rbr}{R_{0}}\right) \times \left( \dfrac{R}{\Rbr} \right)^{\delta_{1}}\ \text{for all } R.
\end{equation}

The primary source injection spectra of all kinds of nuclei are assumed to be a broken power law form independently. For Scheme I and II, each of them includes a low-rigidity break and a high-rigidity break , which is represented as:
\begin{equation}
  q_{\mathrm{i}} \propto  N_{i} \times \left\{ \begin{array}{ll}
                                                          \left( \dfrac{R}{R\mathrm{_{1}^{i}}} \right)^{-\nu_{1}^{i}} & R \leq R_{1}^{i}\\
                                                          \left( \dfrac{R}{R\mathrm{_{1}^{i}}} \right)^{-\nu_{2}^{i}} & R_{1}^{i} < R \leq R_{2}^{i} \\
                                                          \left( \dfrac{R}{R\mathrm{_{2}^{i}}} \right)^{-\nu_{3}^{i}} \left( \dfrac{R\mathrm{_{2}^{i}}}{R\mathrm{_{1}^{i}}} \right)^{-\nu_{2}^{i}} & R > R_{2}^{i} 
  \end{array}
  \right.,
  \label{eq:injection_spectra_1}
\end{equation}
where $i$ denotes the species of nuclei, $N_{i}$ is the relative abundance of the species $i$ to that of proton\footnote{The relative abundance of proton is fixed to $10^6$ and the post-propagated normalization flux of protons at 100 GeV is fixed to $4.45 \times 10^{-2}\m^{-2}\s^{-1}\sr^{-1}\GeV^{-1}$.}, and $\nu \equiv \nu_{1}^{i}(\nu_{2}^{i}, \nu_{3}^{i})$ is the spectral index at rigidity $R$ belonging to the several intervals divided by the breaks at the reference rigidity $R_{1}^{i}$ and $R_{2}^{i}$. 
For Scheme III, each of the primary source injection spectra includes only a low-rigidity break:
\begin{equation}
  q_{\mathrm{i}} \propto  N_{i} \times \left\{ \begin{array}{ll}
\left( \dfrac{R}{R\mathrm{_{1}^{i}}} \right)^{-\nu_{1}^{i}} & R \leq R_{1}^{i}\\
\left( \dfrac{R}{R\mathrm{_{1}^{i}}} \right)^{-\nu_{2}^{i}} & R > R_{1}^{i}  
                                               \end{array}
                                             \right..
\label{eq:injection_spectra_2}
\end{equation}

In this work, we use independent primary source injection spectra for He, C, N, O, Ne, Mg, and Si. \footnote{Here, we use the injection spectra of the dominating isotopes $^{04}_{02}\mathrm{He}$, $^{12}_{06}\mathrm{C}$, $^{14}_{07}\mathrm{N}$, $^{16}_{08}\mathrm{O}$, $^{20}_{10}\mathrm{Ne}$, $^{24}_{12}\mathrm{Mg}$, and $^{28}_{14}\mathrm{Si}$ to represent that of the corresponding elements. All the other primary injection species who have small contributions on the flux of B are assumed to have the same injection spectra as $^{20}_{10}\mathrm{Ne}$.}, The nuclear network used in our calculations is extended to silicon-28.

The force-field approximation \citep{Gleeson1968} is adopted to describe the effects of solar modulation in the solar system, which contains only one parameter the so-called solar-modulation potential $\phi$. All the above configurations are simulated and the diffusion equation are solved by the public code {\sc galprop} v56 \footnote{http://galprop.stanford.edu} \citep{Strong1998,Moskalenko2002,Strong2001,Moskalenko2003,Ptuskin2006} numerically.\footnote{More details about the configuration can be referred to in \citet{Niu2018,Niu2019}.}

It is necessary to note that in the model described above for Scheme I, the hardening in the spectra at a few hundred GV seems to be repeatedly contributed by the primary source acceleration ($R_{1}^{i},  R_{2}^{i}, \nu_{1}^{i}, \nu_{2}^{i}, \nu_{3}^{i}$) and propagation process ($\Rbr, \delta_{1}, \delta_{2}$). But the former will lead to an equal hardening of the primary and secondary spectra, while the latter will lead to a larger hardening in secondary spectra than in primary ones. The fact whether the secondary nuclei spectra harden even more than that of the primary ones can be directly tested by comparing the differences between $\delta_{1}$ and $\delta_{2}$.

\subsection{Fitting Procedure}
\label{sec:fitting_pro}

 In this work, the Bayesian inference is used to get the posterior probability distribution function (PDF), which is based on the following formula

\begin{equation}
 \label{eq:Bayes}
 p(\boldsymbol{\theta}|D) \propto \mathcal{L}(D|\boldsymbol\theta)\pi(\boldsymbol\theta),
\end{equation}
where $\boldsymbol{\theta}=\{\theta_{1},\dots,\theta_{m}\}$ is the free parameter set, $D$ is the experimental data set, $\mathcal{L}(D|\boldsymbol\theta)$ is the likelihood function, and $\pi(\boldsymbol\theta)$ is the prior PDF which represents our state of knowledge on the values of the parameters before taking into account of the new data. 

 We take the prior PDF as uniform distributions
 \begin{equation}
 \label{eq:priors}
\pi(\theta_{i}) \propto
\left\{
\begin{tabular}{ll}
1, &  \text{for } $\theta_{i,\text{min}}<\theta_{i}<\theta_{i,\text{max}}$
\\
0, & \text{otherwise}
\end{tabular}
\right. 
,
\end{equation}
 and the likelihood function as a Gaussian form
 \begin{equation}
 \label{eq:likelihood}
\mathcal{L}(D|\boldsymbol\theta)=
\prod_{i}
\frac{1}{\sqrt{2\pi \sigma_{i}^{2}}}
\exp\left[
  -\frac{(f_{\text{th},i}(\boldsymbol\theta)-f_{\text{exp},i})^{2}}{2\sigma_{i}^{2}}  
\right]  ,  
\end{equation}
where $f_{\text{th},i}(\boldsymbol\theta)$ is the predicted $i$-th observable from the model which depends on the parameter set $\boldsymbol\theta$, and $f_{\text{exp},i}$ is the one measured by the experiment with uncertainty $\sigma_{i}$.

Here we use the Markov Chain Monte Carlo (MCMC) algorithms which is proposed by \citet{Goodman2010} instead of classical Metropolis-Hastings to determine the PDFs of the parameters, because its ensemble samplers can avoid the Markov Chains falls into local optimal values and thus provide us robust PDFs of the parameters. The algorithm proposed by \citet{Goodman2010} is slightly altered and implemented as the {\tt Python} module {\tt emcee}\footnote{http://dan.iel.fm/emcee/} by \citet{emcee}, which makes it easy to use by the advantages of {\tt Python}.

In total, for Scheme I, we have the following 50 free parameters:
\begin{align*}
  \boldsymbol{\theta}_{\mathrm{I}} =  &\{ D_{0}, \eta, \Rbr, \delta_{1}, \delta_{2}, z_{h}, v_{A}, \phi, | \\
                         &N_{\mathrm{He}}, R_{1}^{\mathrm{He}},  R_{2}^{\mathrm{He}}, \nu_{1}^{\mathrm{He}}, \nu_{2}^{\mathrm{He}}, \nu_{3}^{\mathrm{He}}, | \\
                         &N_{\mathrm{C}}, R_{1}^{\mathrm{C}},  R_{2}^{\mathrm{C}}, \nu_{1}^{\mathrm{C}}, \nu_{2}^{\mathrm{C}}, \nu_{3}^{\mathrm{C}}, | \\
                         &N_{\mathrm{N}},  R_{1}^{\mathrm{N}},  R_{2}^{\mathrm{N}}, \nu_{1}^{\mathrm{N}}, \nu_{2}^{\mathrm{N}}, \nu_{3}^{\mathrm{N}}, | \\
                         &N_{\mathrm{O}},  R_{1}^{\mathrm{O}},  R_{2}^{\mathrm{O}}, \nu_{1}^{\mathrm{O}}, \nu_{2}^{\mathrm{O}}, \nu_{3}^{\mathrm{O}}, | \\
                         &N_{\mathrm{Ne}},  R_{1}^{\mathrm{Ne}},  R_{2}^{\mathrm{Ne}}, \nu_{1}^{\mathrm{Ne}}, \nu_{2}^{\mathrm{Ne}}, \nu_{3}^{\mathrm{Ne}}, | \\
                         &N_{\mathrm{Mg}},  R_{1}^{\mathrm{Mg}},  R_{2}^{\mathrm{Mg}}, \nu_{1}^{\mathrm{Mg}}, \nu_{2}^{\mathrm{Mg}}, \nu_{3}^{\mathrm{Mg}}, | \\
                         &N_{\mathrm{Si}},  R_{1}^{\mathrm{Si}},  R_{2}^{\mathrm{Si}}, \nu_{1}^{\mathrm{Si}}, \nu_{2}^{\mathrm{Si}}, \nu_{3}^{\mathrm{Si}} \}~.
\end{align*}

For Scheme II, we have the following 48 free parameters:
\begin{align*}
  \boldsymbol{\theta}_{\mathrm{II}} =  &\{ D_{0}, \eta, \delta_{1}, z_{h}, v_{A}, \phi, | \\
                         &N_{\mathrm{He}}, R_{1}^{\mathrm{He}},  R_{2}^{\mathrm{He}}, \nu_{1}^{\mathrm{He}}, \nu_{2}^{\mathrm{He}}, \nu_{3}^{\mathrm{He}}, | \\
                         &N_{\mathrm{C}}, R_{1}^{\mathrm{C}},  R_{2}^{\mathrm{C}}, \nu_{1}^{\mathrm{C}}, \nu_{2}^{\mathrm{C}}, \nu_{3}^{\mathrm{C}}, | \\
                         &N_{\mathrm{N}},  R_{1}^{\mathrm{N}},  R_{2}^{\mathrm{N}}, \nu_{1}^{\mathrm{N}}, \nu_{2}^{\mathrm{N}}, \nu_{3}^{\mathrm{N}}, | \\
                         &N_{\mathrm{O}},  R_{1}^{\mathrm{O}},  R_{2}^{\mathrm{O}}, \nu_{1}^{\mathrm{O}}, \nu_{2}^{\mathrm{O}}, \nu_{3}^{\mathrm{O}}, | \\
                         &N_{\mathrm{Ne}},  R_{1}^{\mathrm{Ne}},  R_{2}^{\mathrm{Ne}}, \nu_{1}^{\mathrm{Ne}}, \nu_{2}^{\mathrm{Ne}}, \nu_{3}^{\mathrm{Ne}}, | \\
                         &N_{\mathrm{Mg}},  R_{1}^{\mathrm{Mg}},  R_{2}^{\mathrm{Mg}}, \nu_{1}^{\mathrm{Mg}}, \nu_{2}^{\mathrm{Mg}}, \nu_{3}^{\mathrm{Mg}}, | \\
                         &N_{\mathrm{Si}},  R_{1}^{\mathrm{Si}},  R_{2}^{\mathrm{Si}}, \nu_{1}^{\mathrm{Si}}, \nu_{2}^{\mathrm{Si}}, \nu_{3}^{\mathrm{Si}} \}~.
\end{align*}

For Scheme III, we have the following 36 free parameters:
\begin{align*}
  \boldsymbol{\theta}_{\mathrm{III}} =  &\{ D_{0}, \eta, \Rbr, \delta_{1}, \delta_{2}, z_{h}, v_{A}, \phi, | \\
                         &N_{\mathrm{He}}, R_{1}^{\mathrm{He}}, \nu_{1}^{\mathrm{He}}, \nu_{2}^{\mathrm{He}}, | \\
                         &N_{\mathrm{C}}, R_{1}^{\mathrm{C}},  \nu_{1}^{\mathrm{C}}, \nu_{2}^{\mathrm{C}}, | \\
                         &N_{\mathrm{N}},  R_{1}^{\mathrm{N}}, \nu_{1}^{\mathrm{N}}, \nu_{2}^{\mathrm{N}}, | \\
                         &N_{\mathrm{O}},  R_{1}^{\mathrm{O}}, \nu_{1}^{\mathrm{O}}, \nu_{2}^{\mathrm{O}}, | \\
                         &N_{\mathrm{Ne}},  R_{1}^{\mathrm{Ne}}, \nu_{1}^{\mathrm{Ne}}, \nu_{2}^{\mathrm{Ne}}, | \\
                         &N_{\mathrm{Mg}},  R_{1}^{\mathrm{Mg}}, \nu_{1}^{\mathrm{Mg}}, \nu_{2}^{\mathrm{Mg}}, | \\
                         &N_{\mathrm{Si}},  R_{1}^{\mathrm{Si}}, \nu_{1}^{\mathrm{Si}}, \nu_{2}^{\mathrm{Si}}  \}~.
\end{align*}

For all the schemes, the spectral data of He, C, N, O, and B is collected from \citet{AMS7years}, that of Ne, Mg, and Si is collected from \citet{AMS02_Ne_Mg_Si}, and the data errors used in our fitting are the quadratic sum of statistical and systematic errors.

\section{Results}

\begin{figure*}[htbp]
  \centering
  \includegraphics[width=0.32\textwidth]{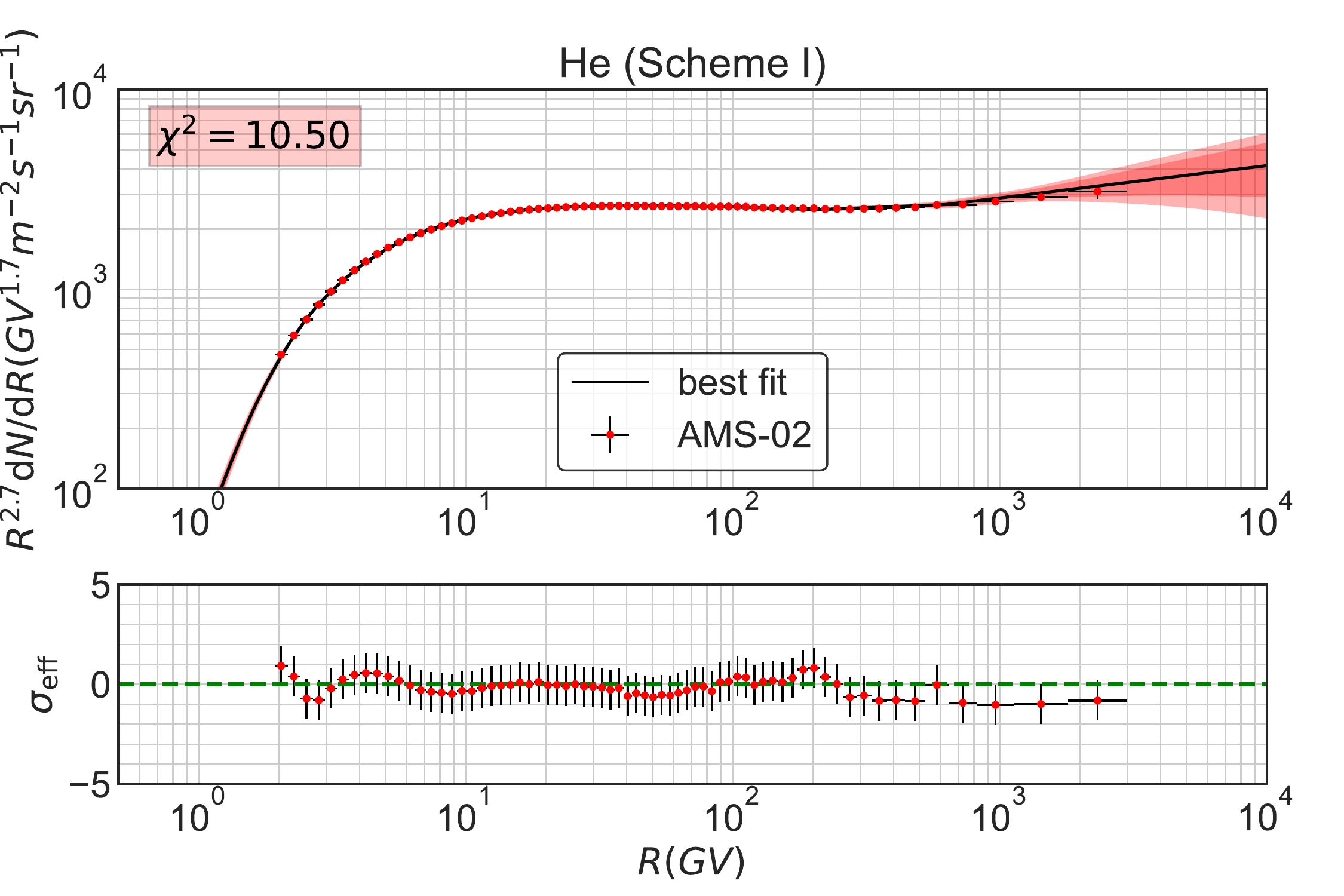}
  \includegraphics[width=0.32\textwidth]{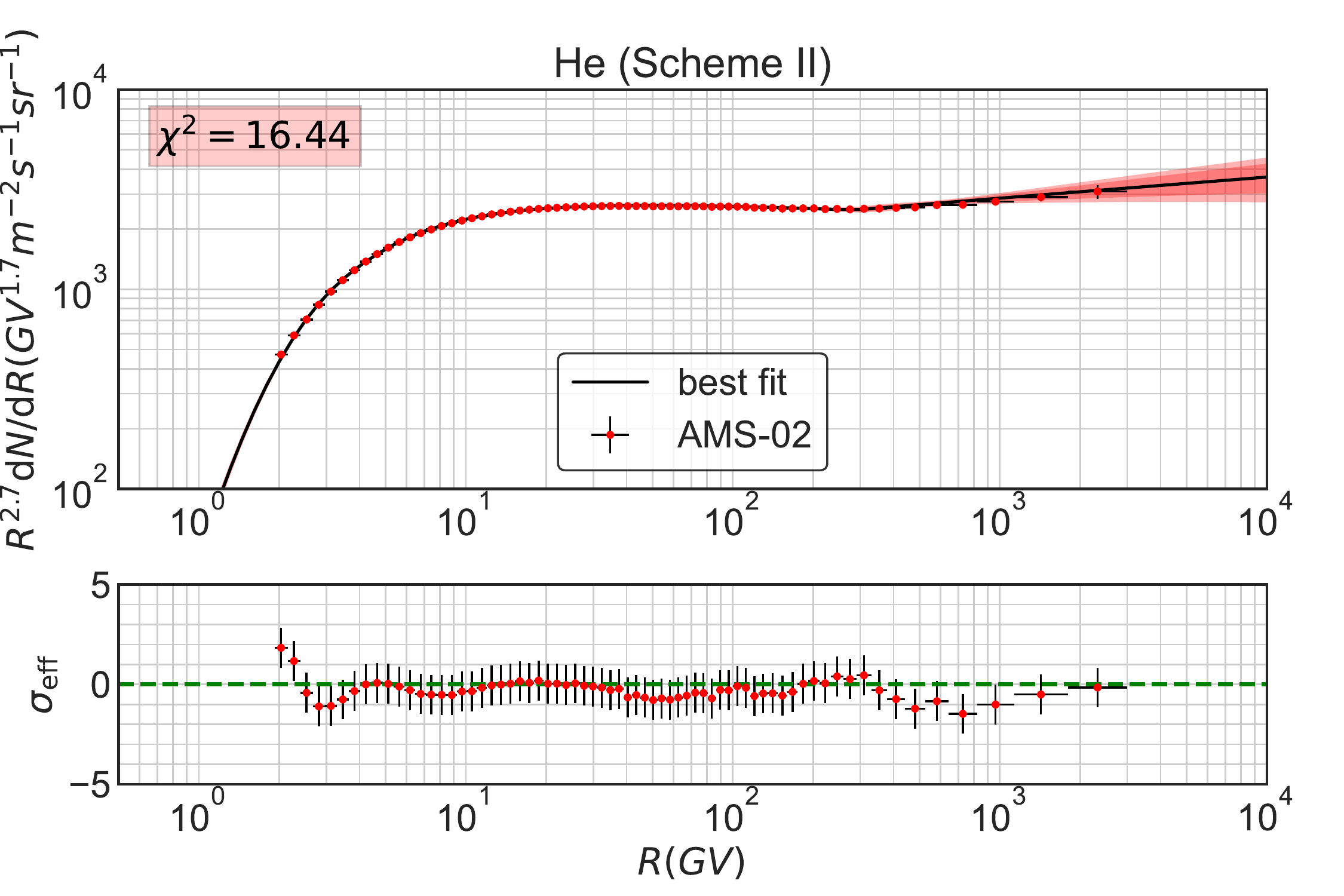}
  \includegraphics[width=0.32\textwidth]{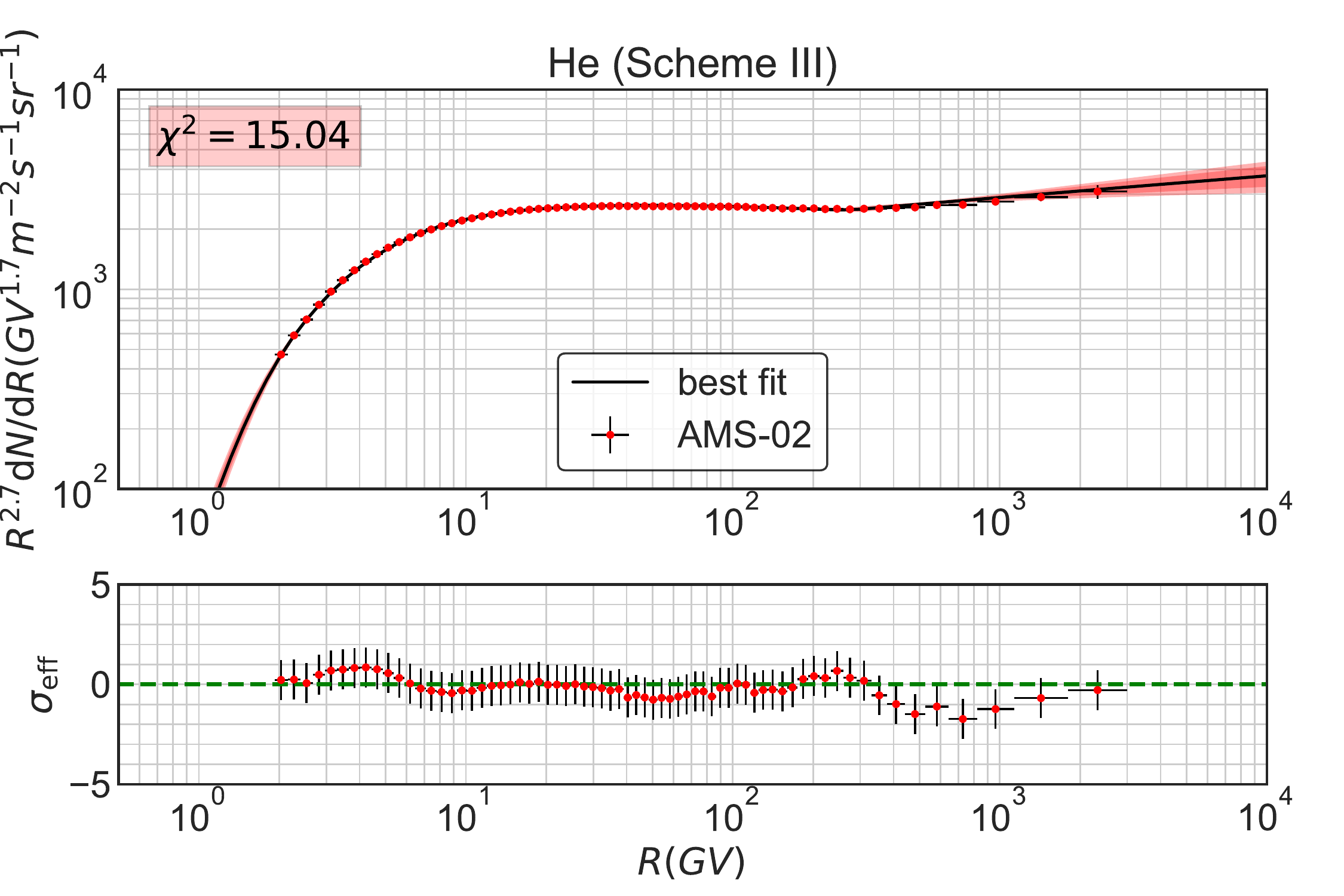}
  \includegraphics[width=0.32\textwidth]{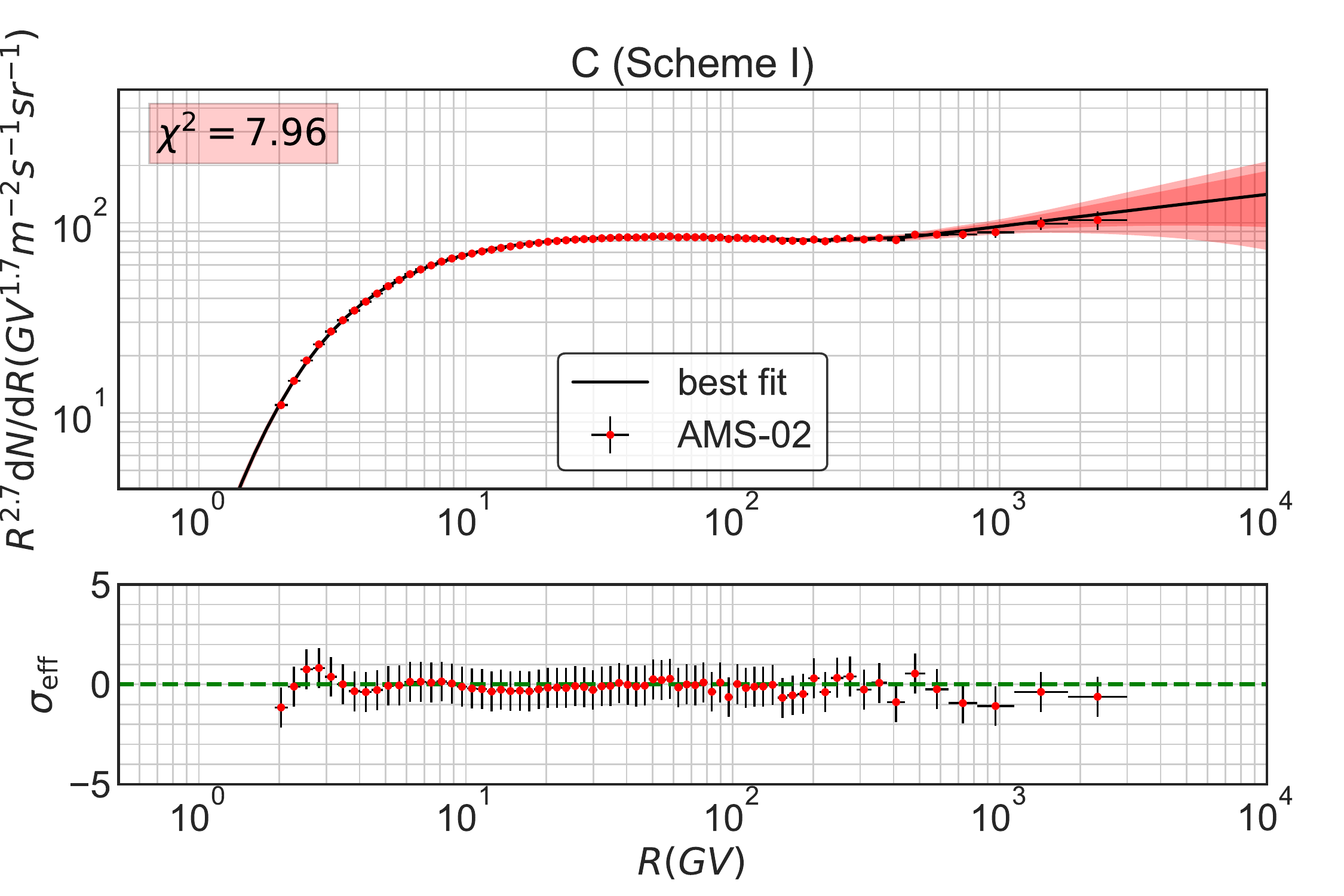}
  \includegraphics[width=0.32\textwidth]{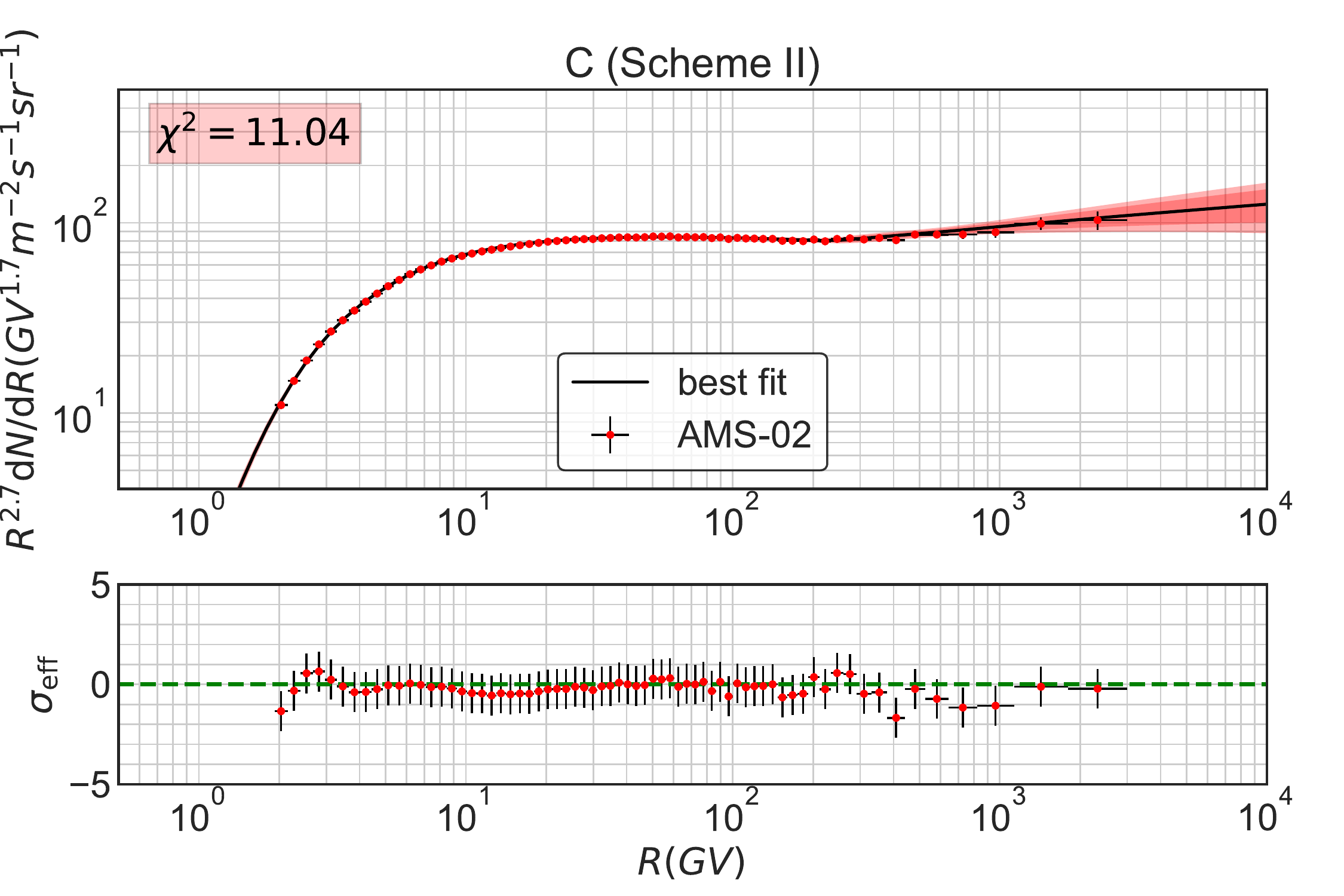}
  \includegraphics[width=0.32\textwidth]{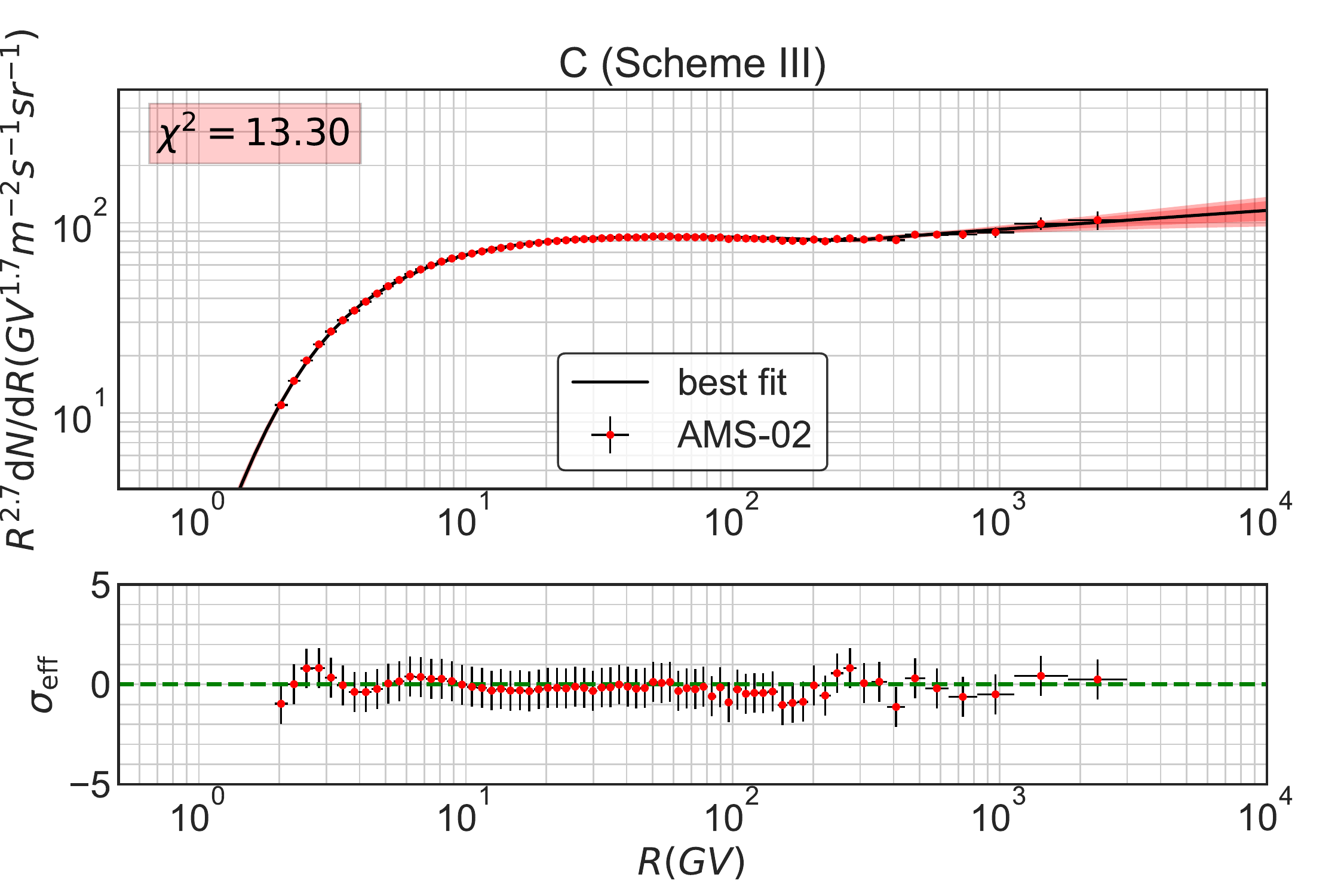}
  \includegraphics[width=0.32\textwidth]{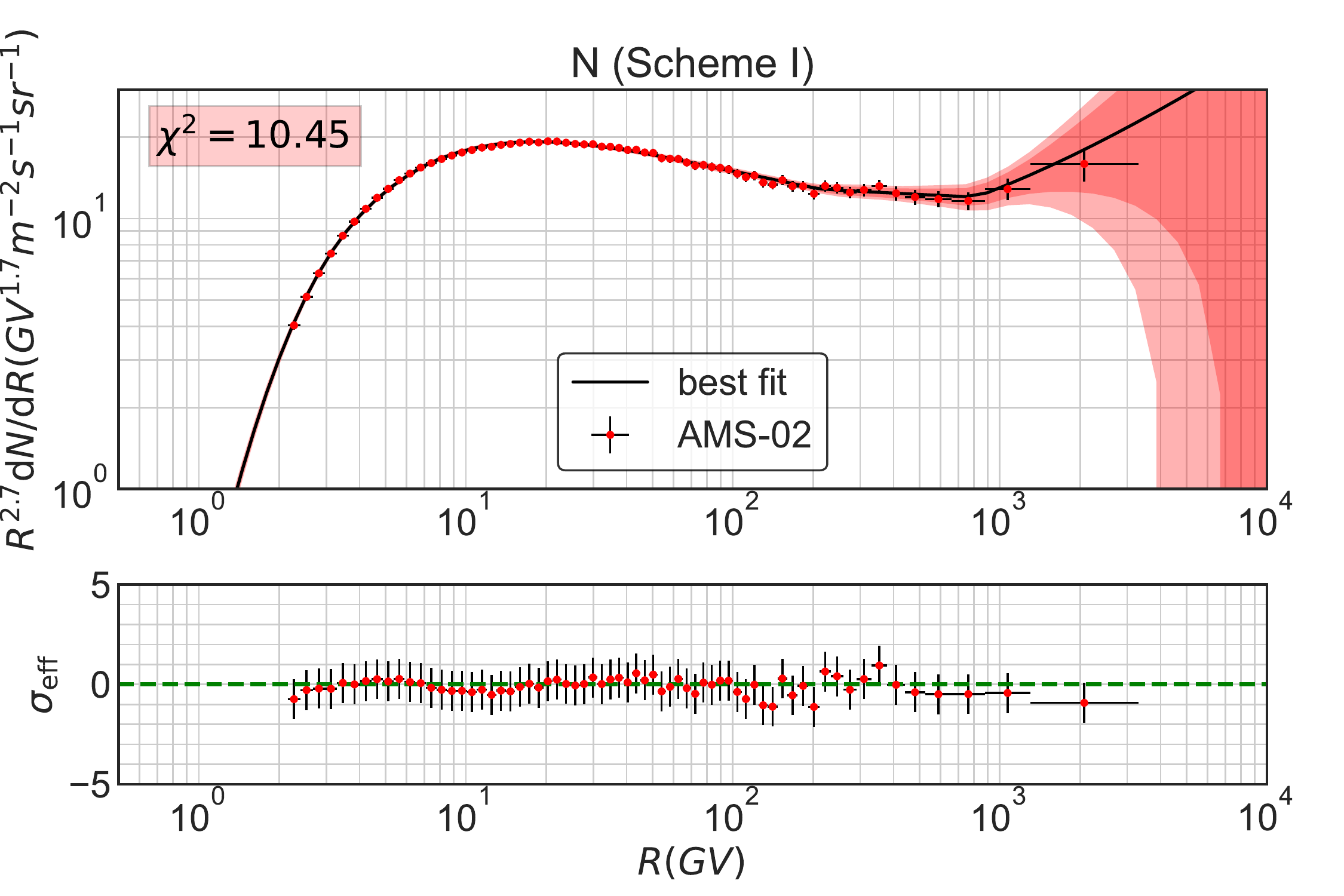}
  \includegraphics[width=0.32\textwidth]{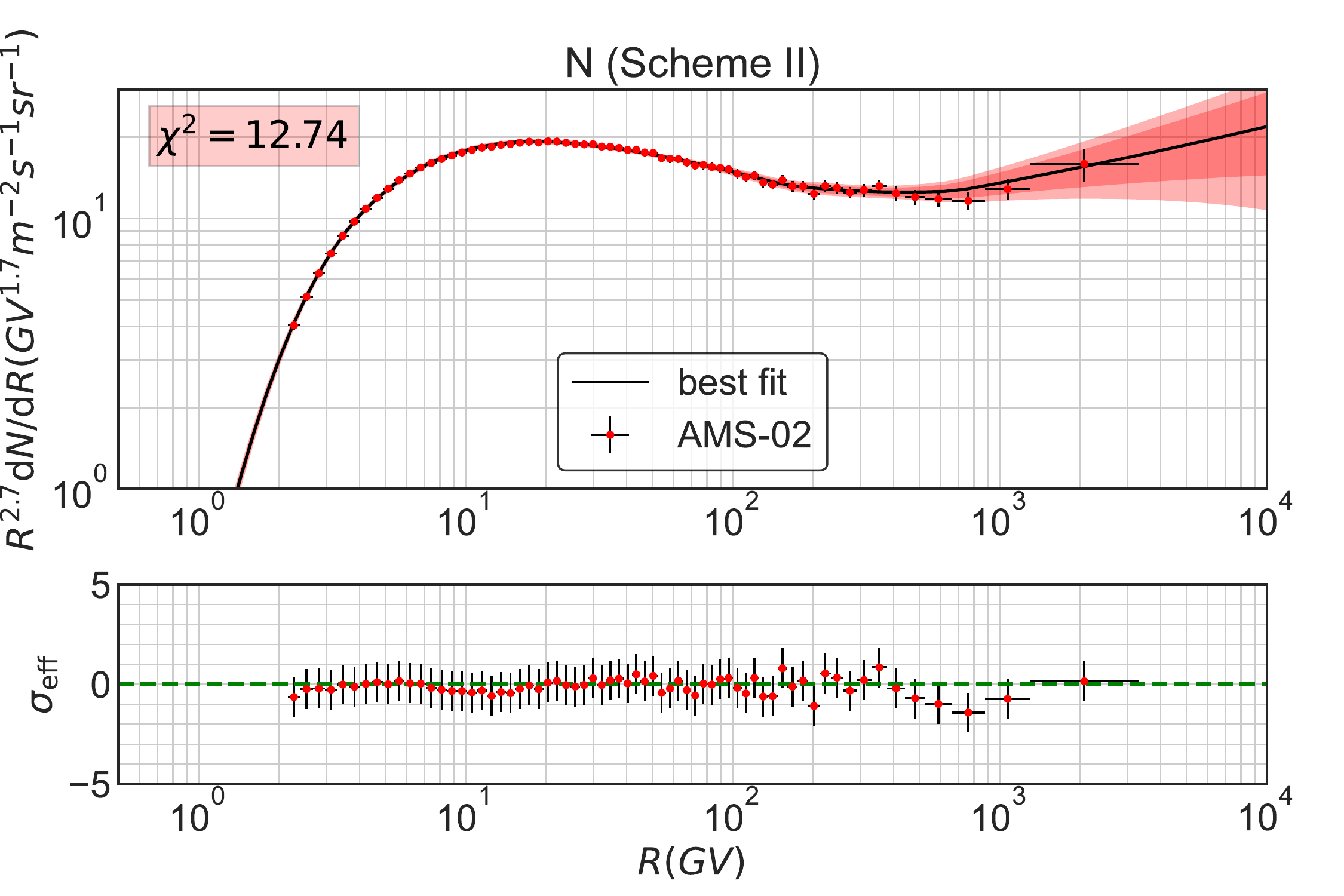}
  \includegraphics[width=0.32\textwidth]{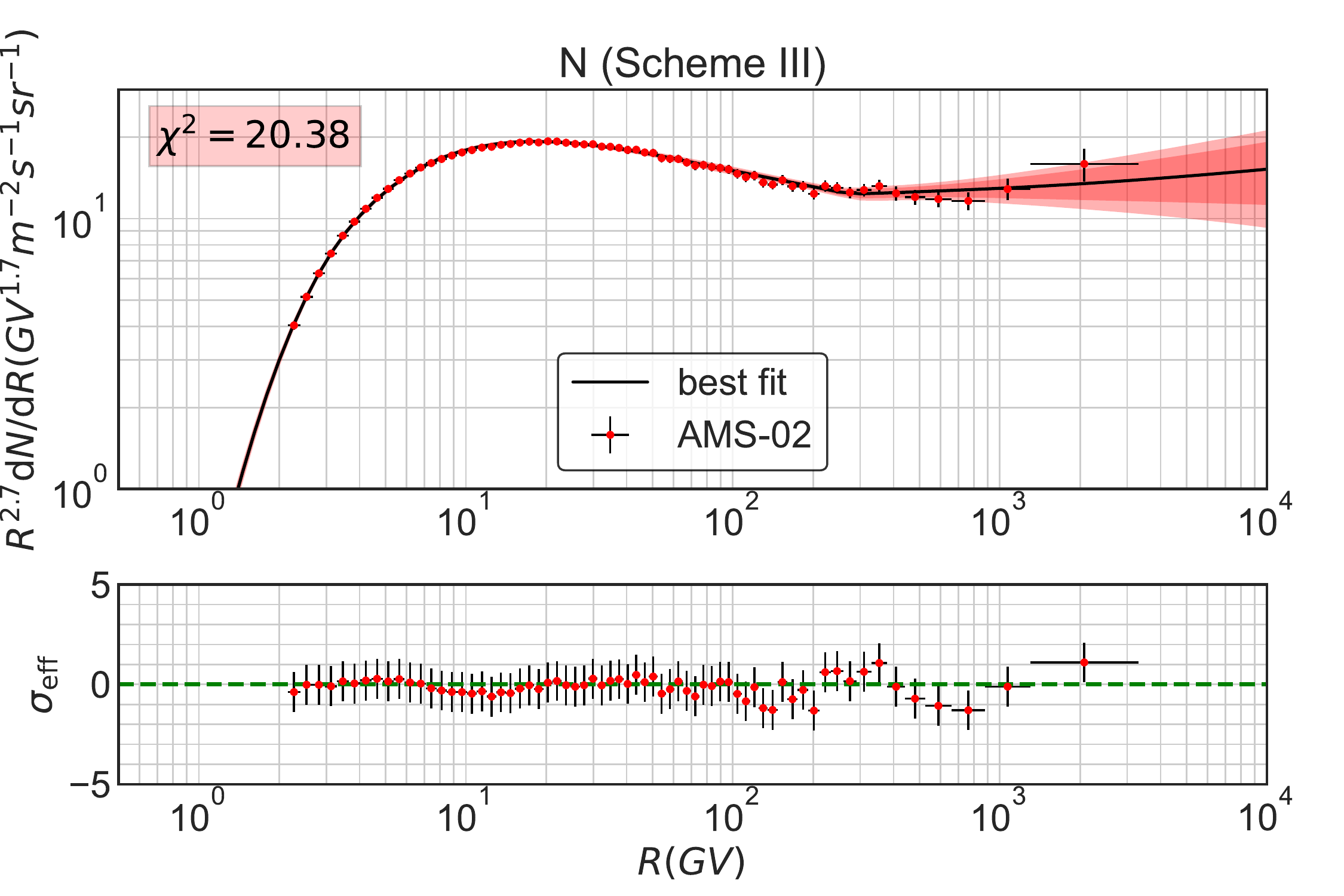}
  \includegraphics[width=0.32\textwidth]{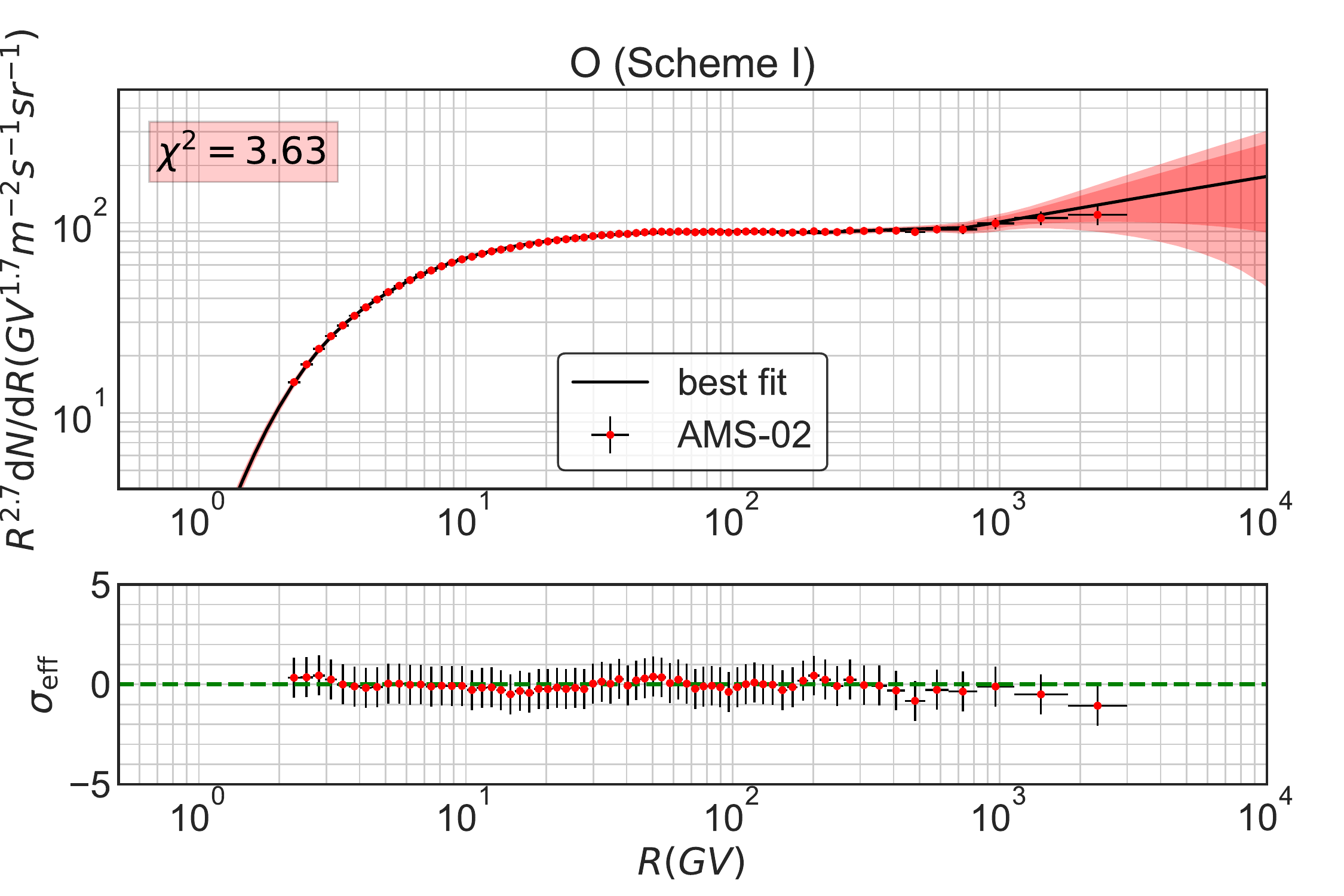}
  \includegraphics[width=0.32\textwidth]{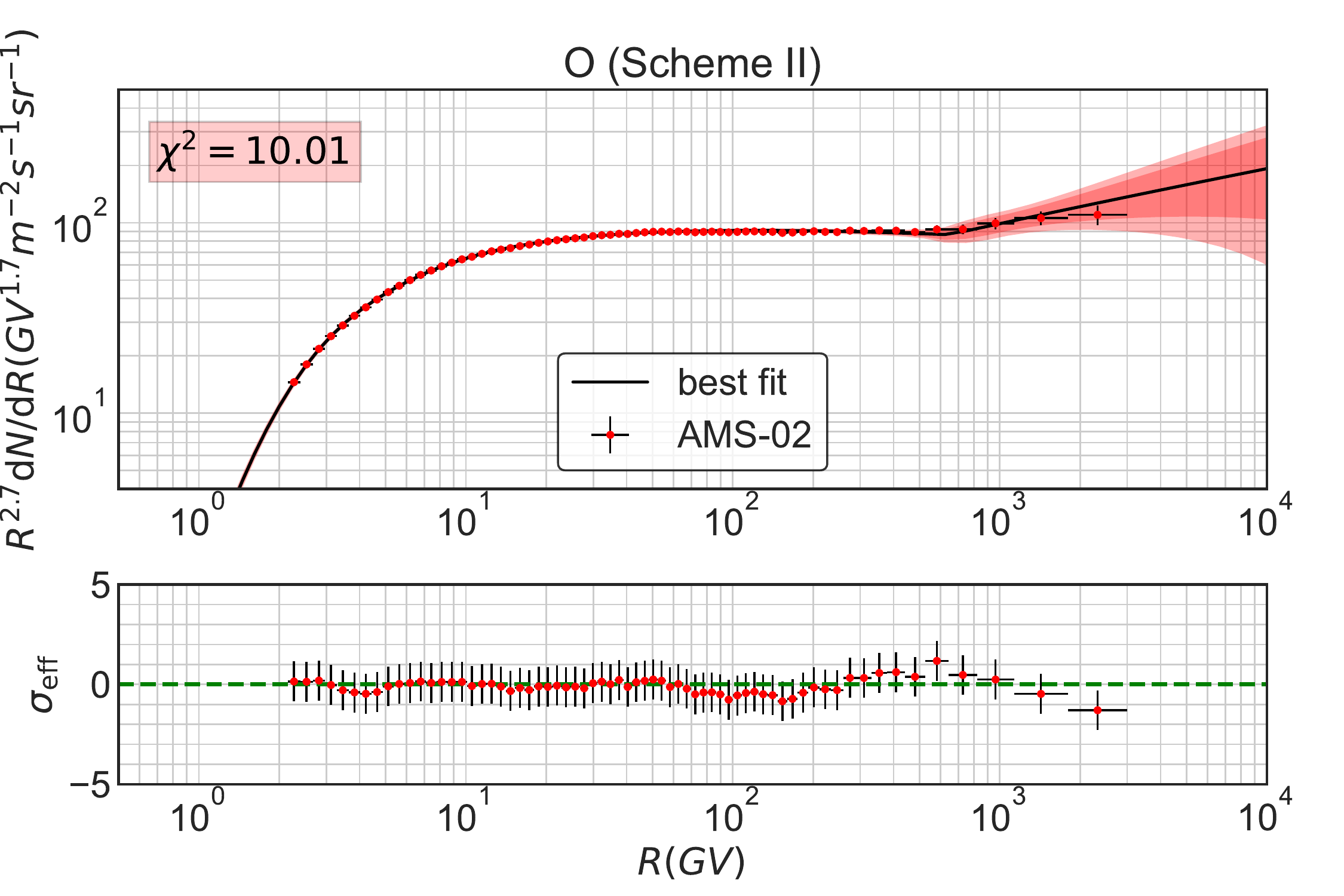}
  \includegraphics[width=0.32\textwidth]{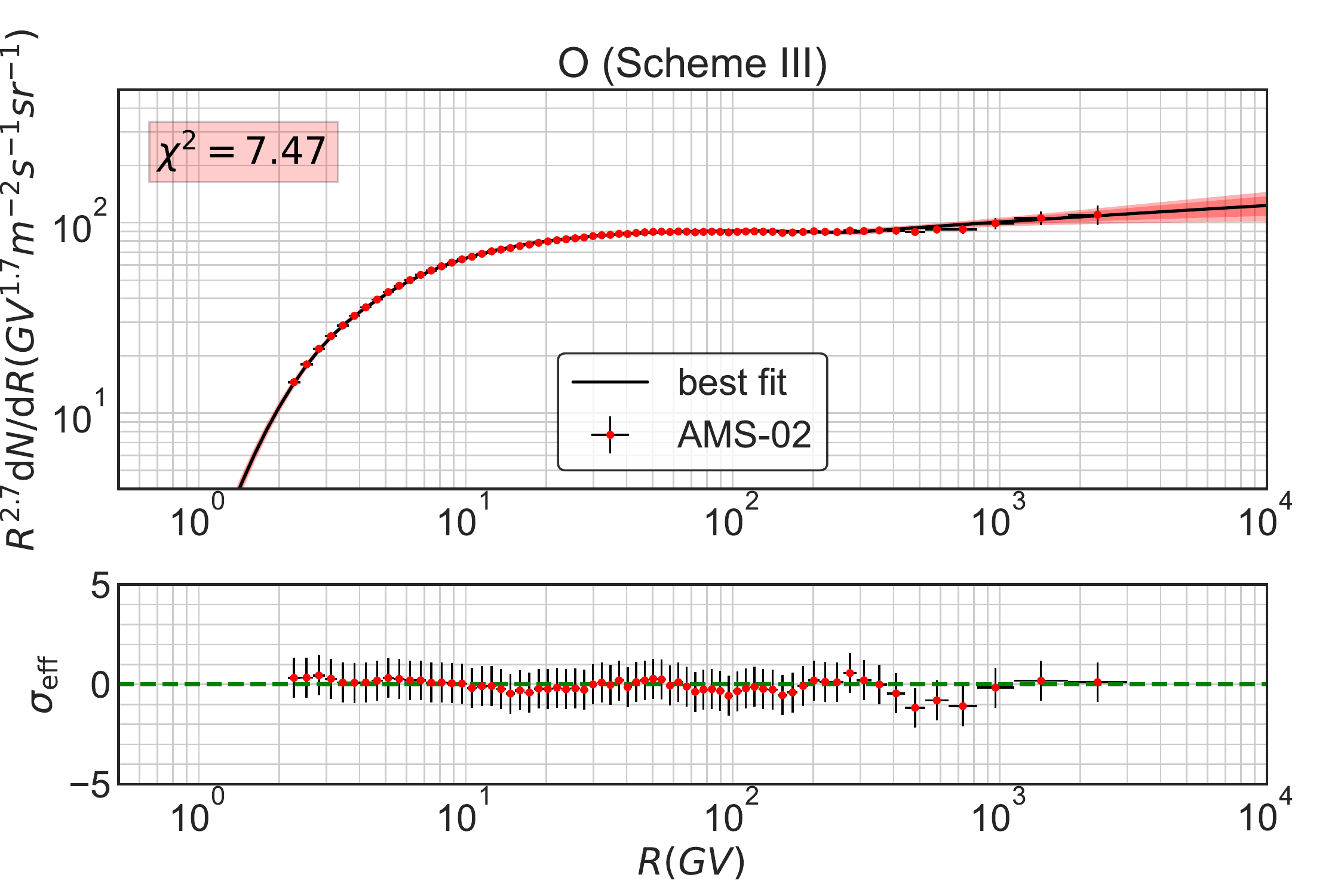}
  \caption{Fitting results and corresponding residuals to the CR spectra of He, C, N, and O for the Scheme I, II, and III. The 2$\sigma$ (deep red) and 3$\sigma$ (light red) bounds are also shown in the subfigures. The relevant $\chi^2$ of each spectrum is given in the subfigures as well.}
  \label{fig:spectra01}
\end{figure*}

\begin{figure*}[htbp]
  \centering
  \includegraphics[width=0.32\textwidth]{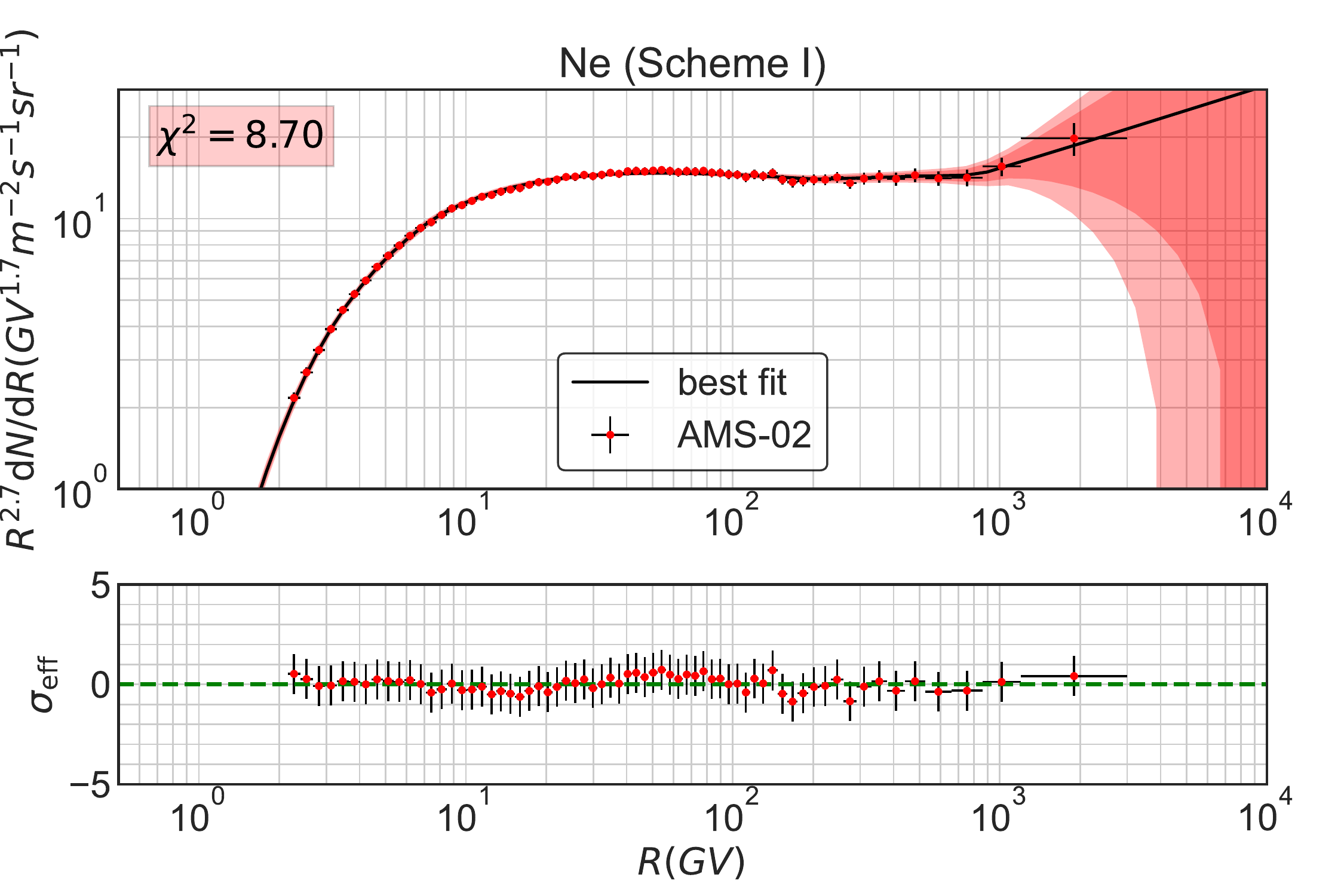}
  \includegraphics[width=0.32\textwidth]{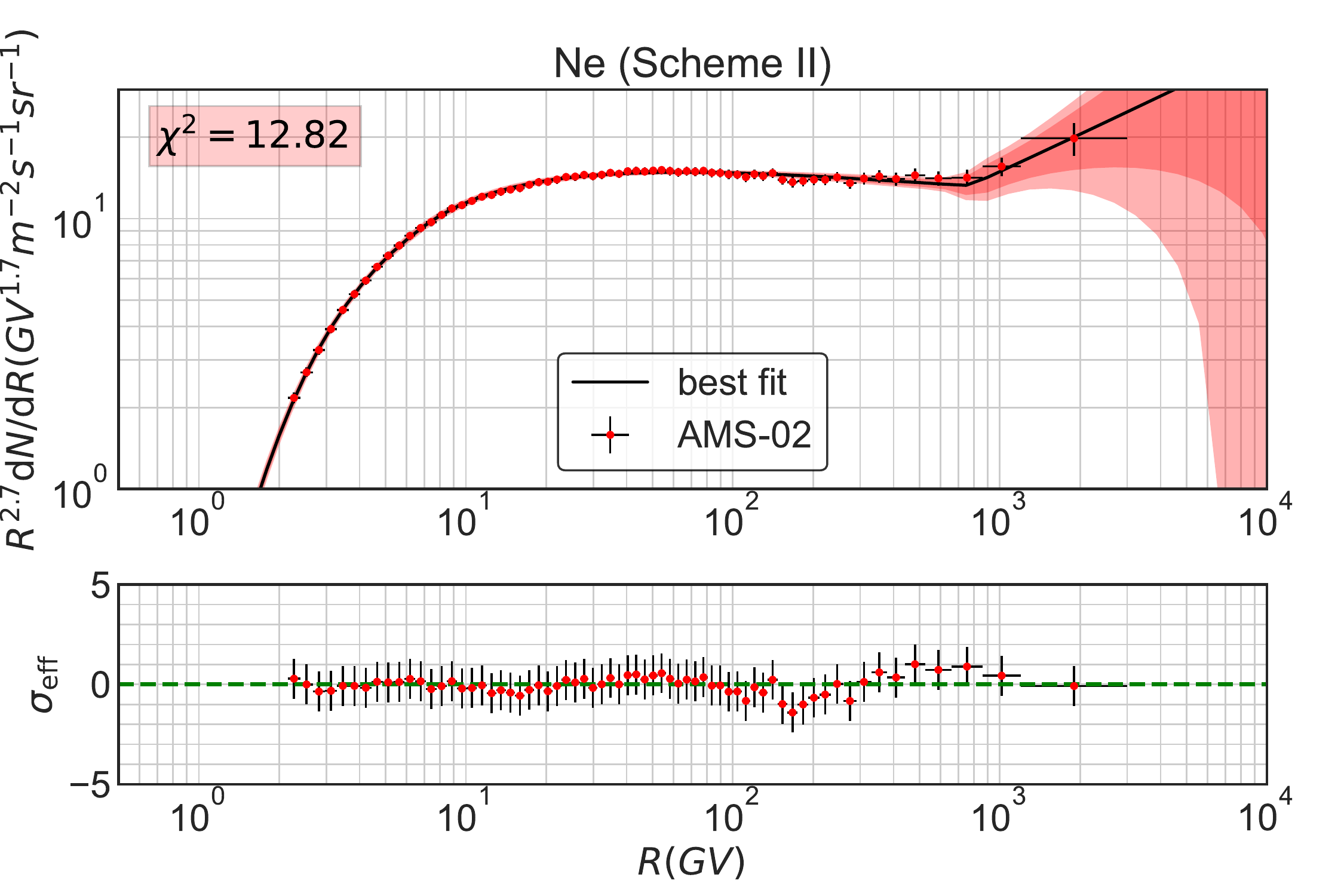}
  \includegraphics[width=0.32\textwidth]{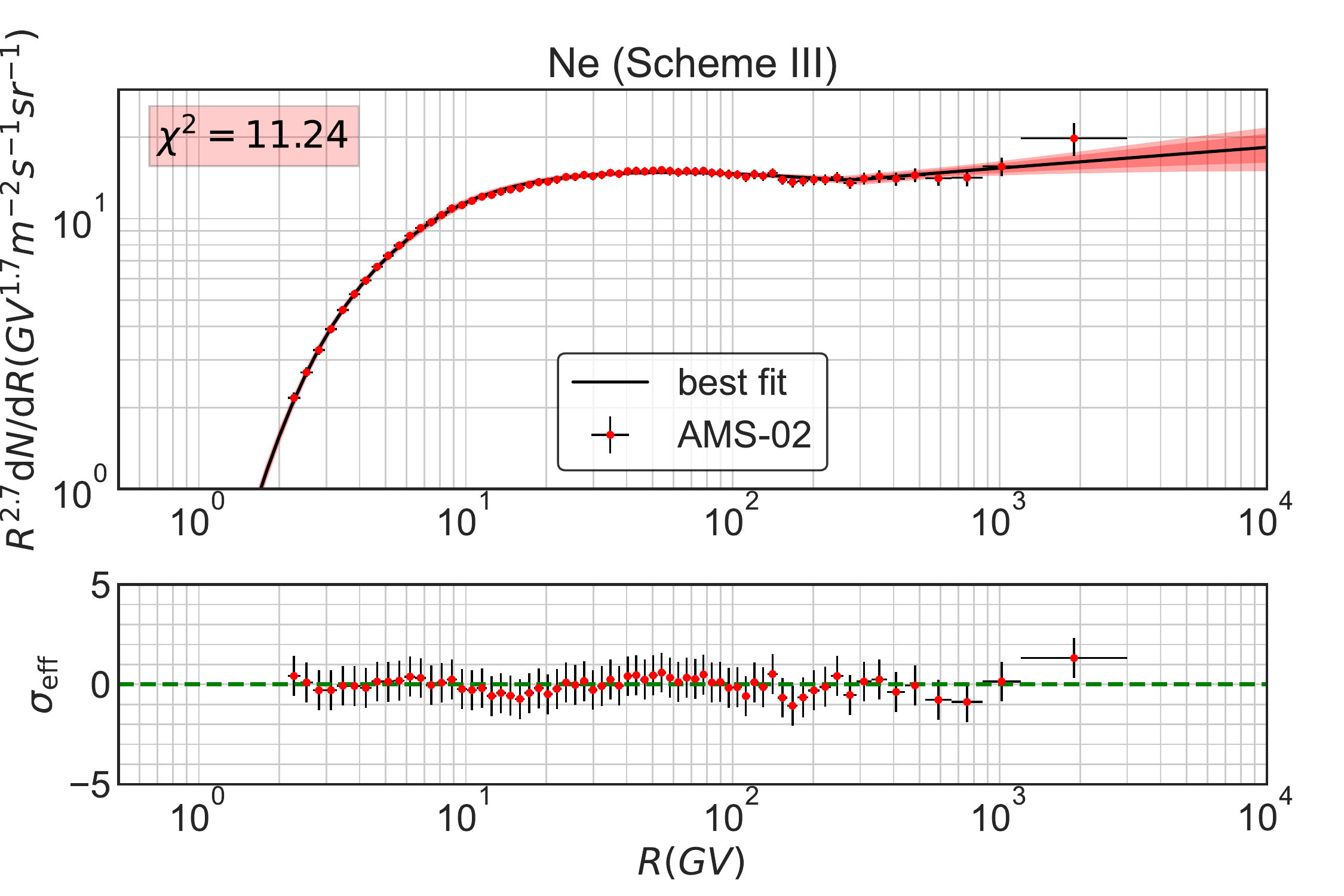}
  \includegraphics[width=0.32\textwidth]{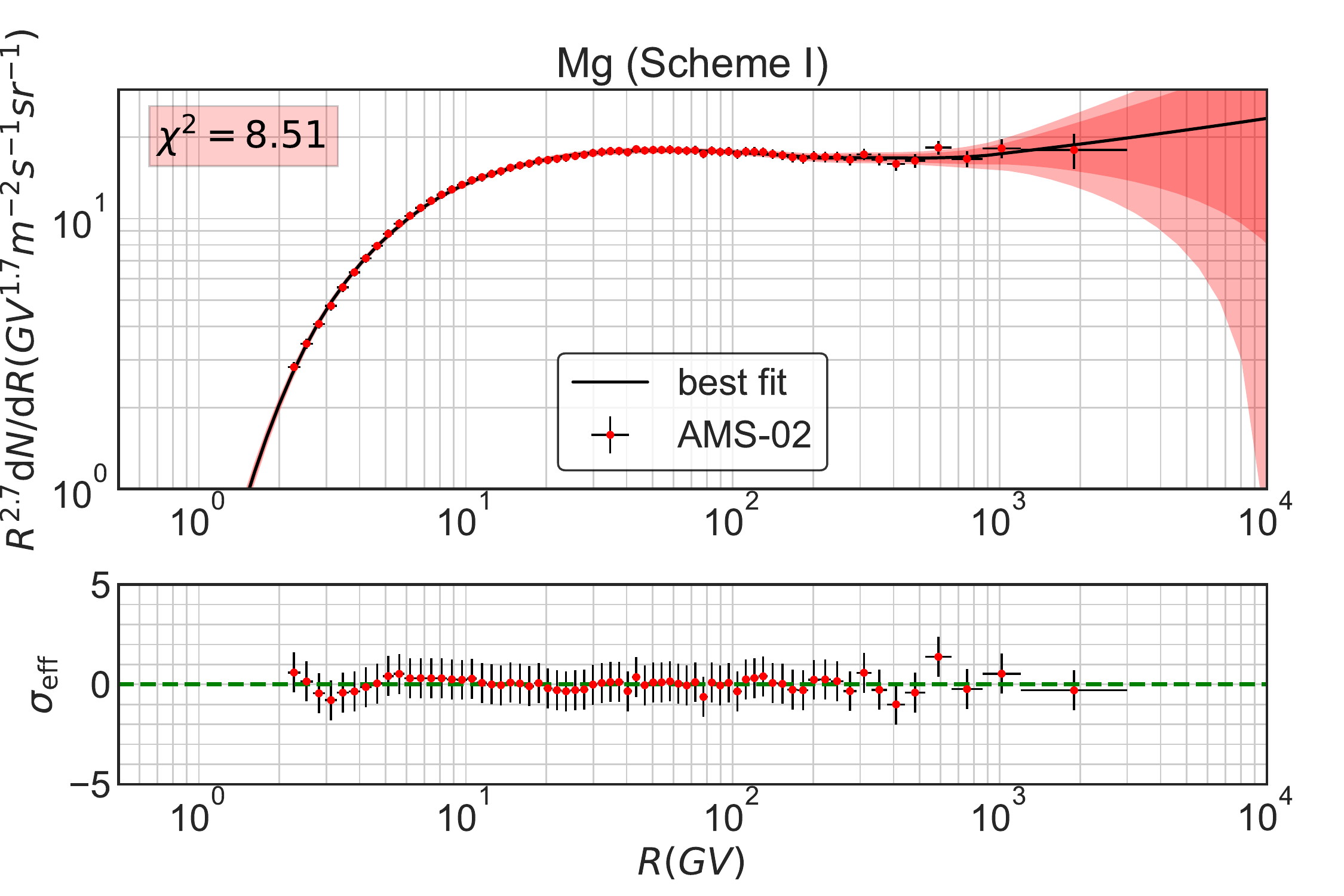}
  \includegraphics[width=0.32\textwidth]{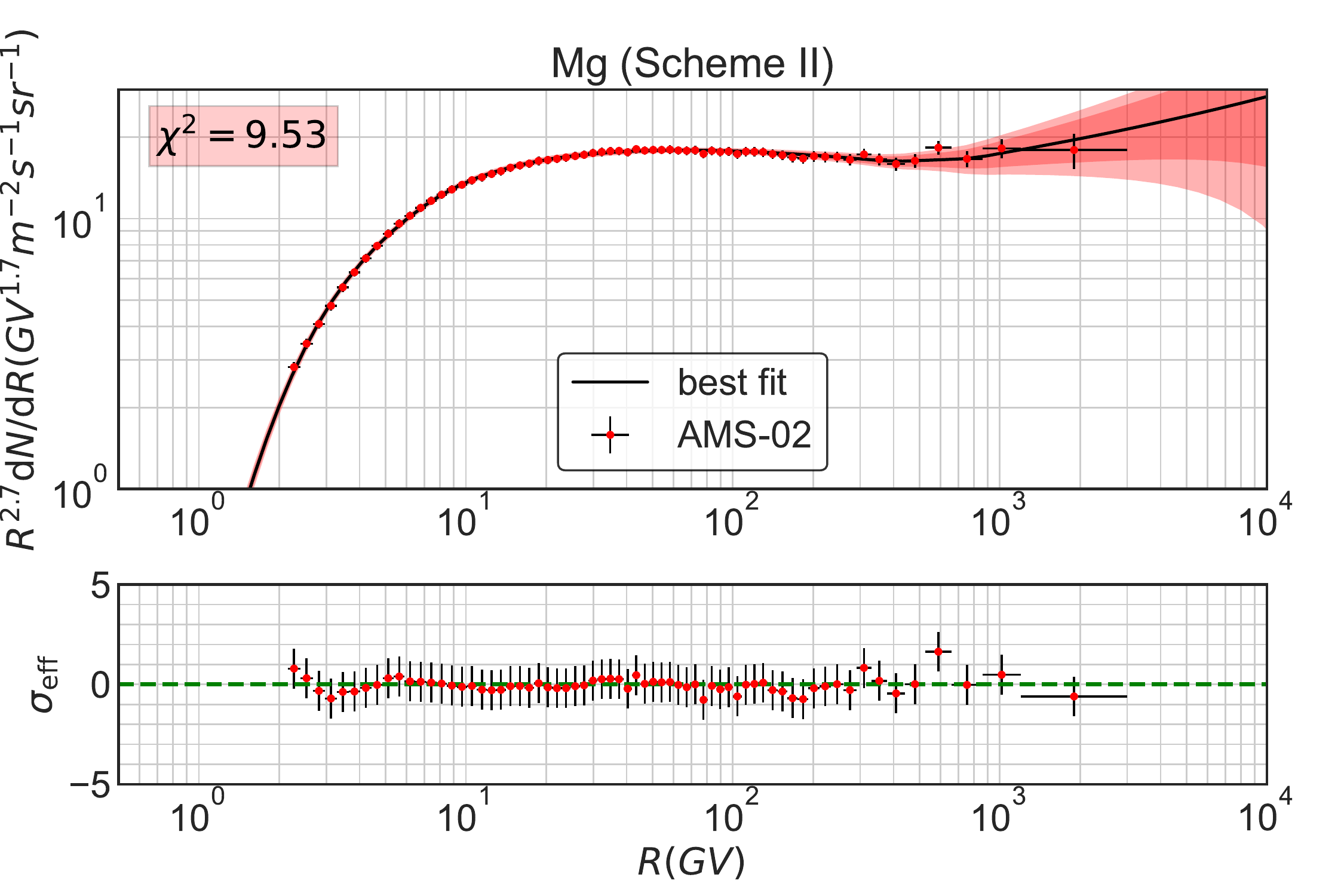}
  \includegraphics[width=0.32\textwidth]{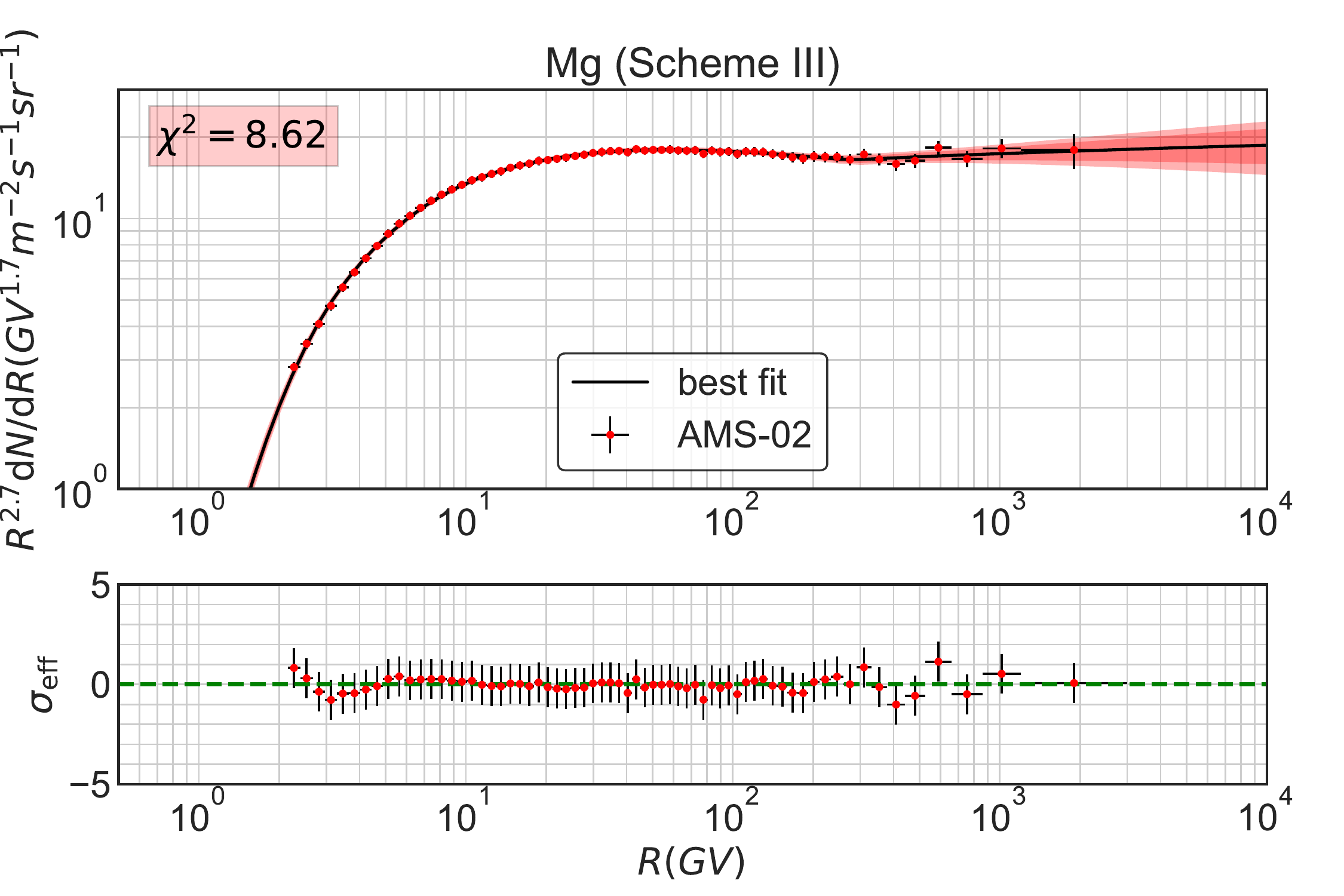}
  \includegraphics[width=0.32\textwidth]{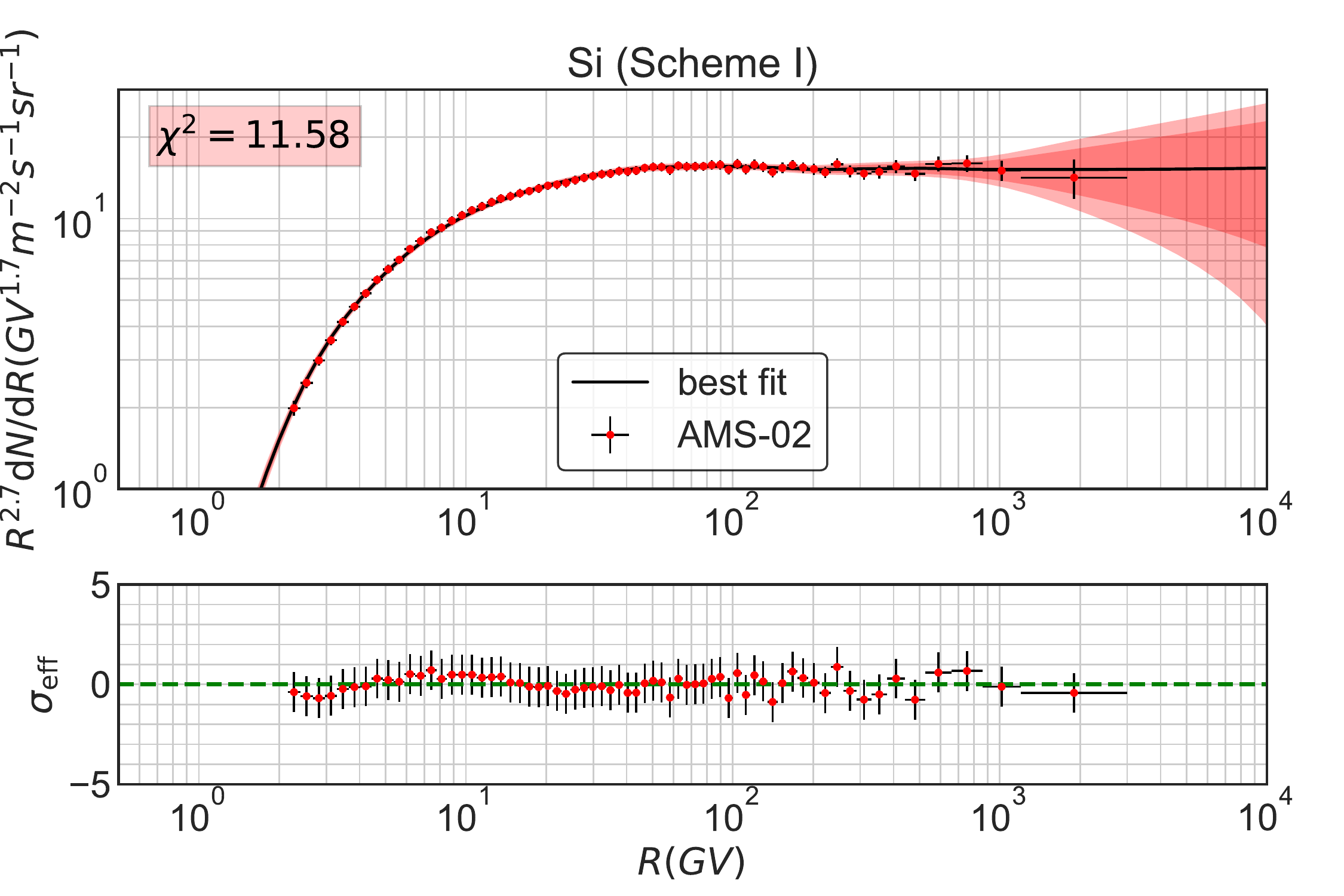}
  \includegraphics[width=0.32\textwidth]{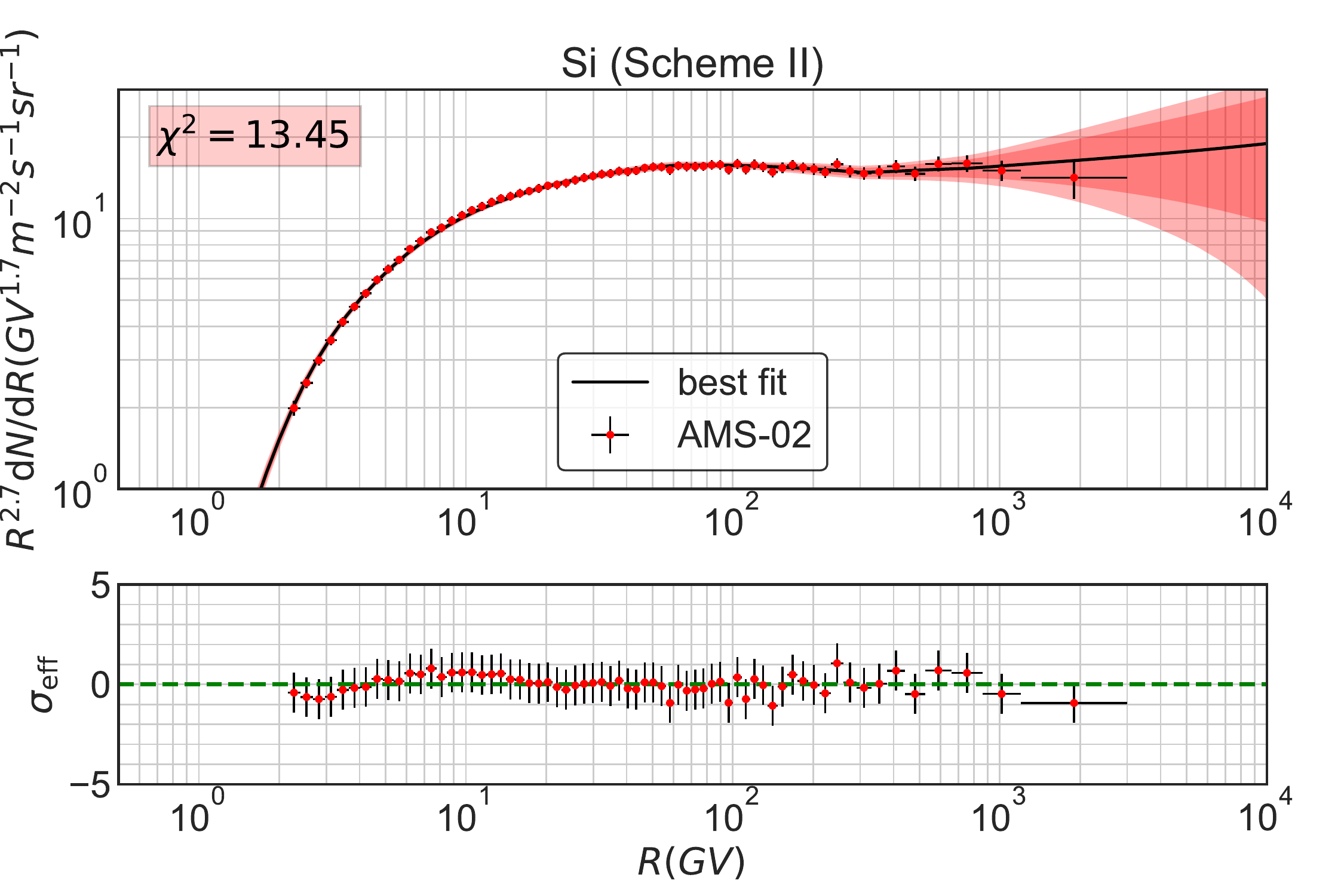}
  \includegraphics[width=0.32\textwidth]{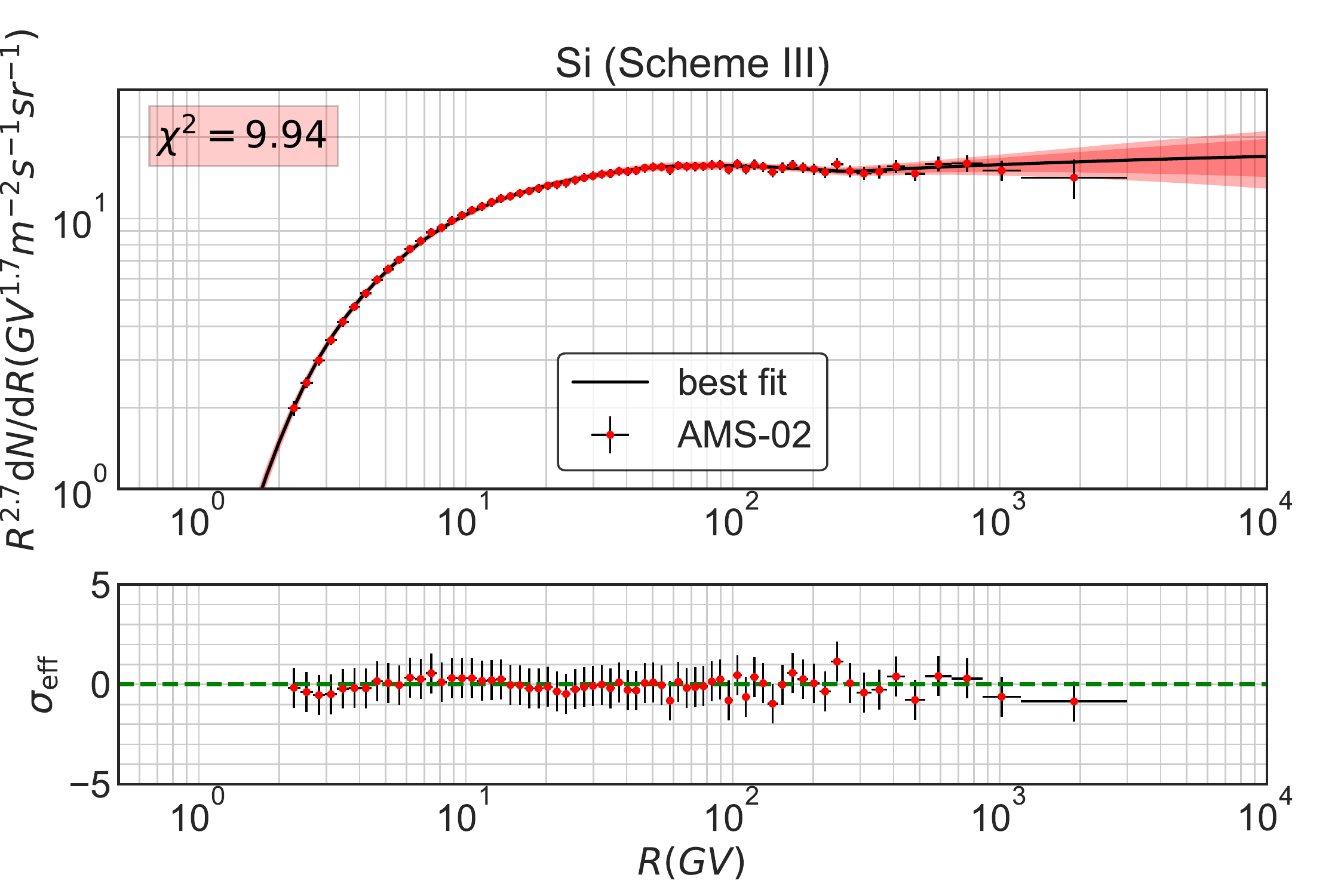}
  \caption{Fitting results and corresponding residuals to the CR spectra of Ne, Mg, and Si for the Scheme I, II, and III. The 2$\sigma$ (deep red) and 3$\sigma$ (light red) bounds are also shown in the subfigures. The relevant $\chi^2$ of each spectrum is given in the subfigures as well.}
  \label{fig:spectra02}
\end{figure*}

\begin{figure*}[htbp]
  \centering
  \includegraphics[width=0.32\textwidth]{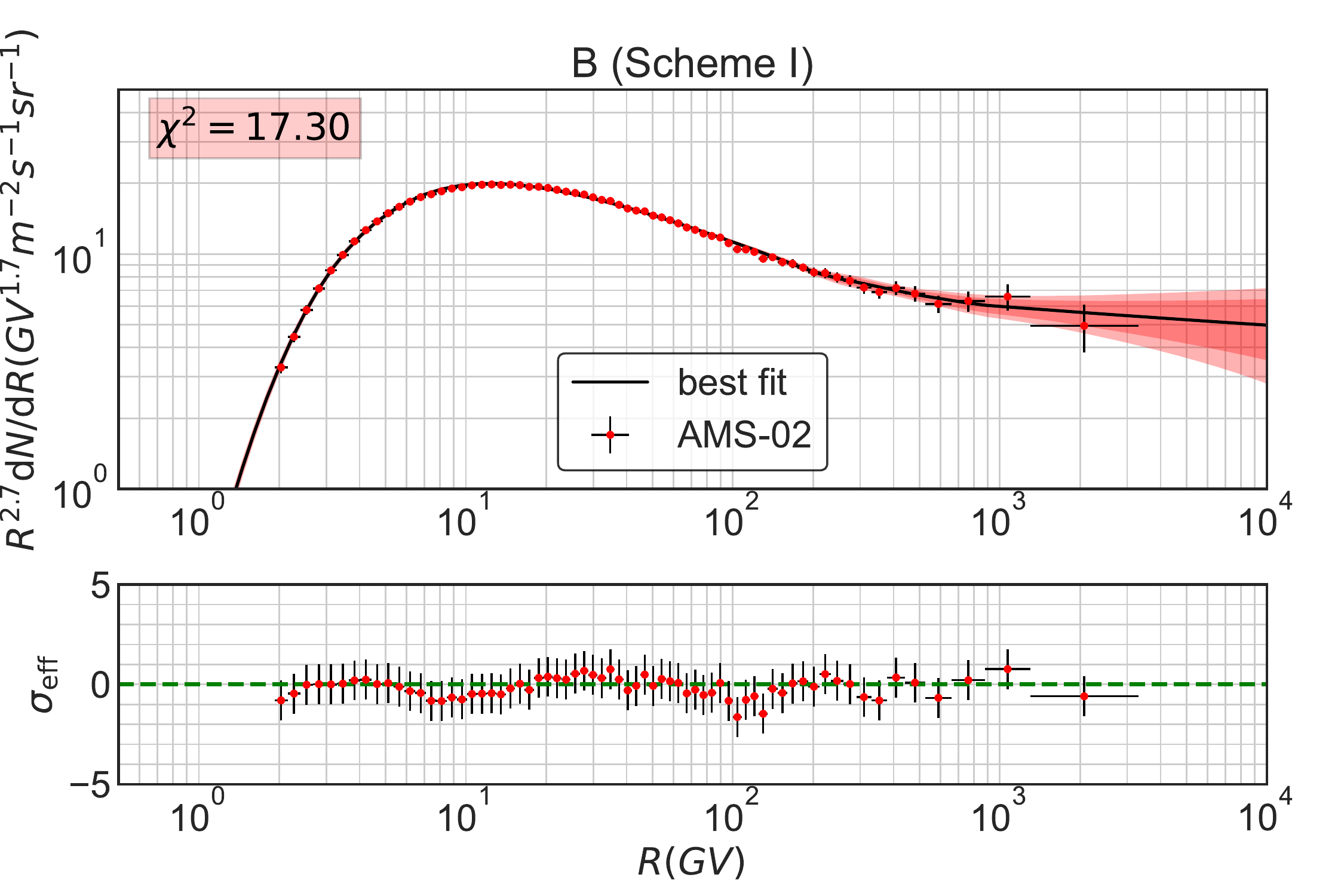}
  \includegraphics[width=0.32\textwidth]{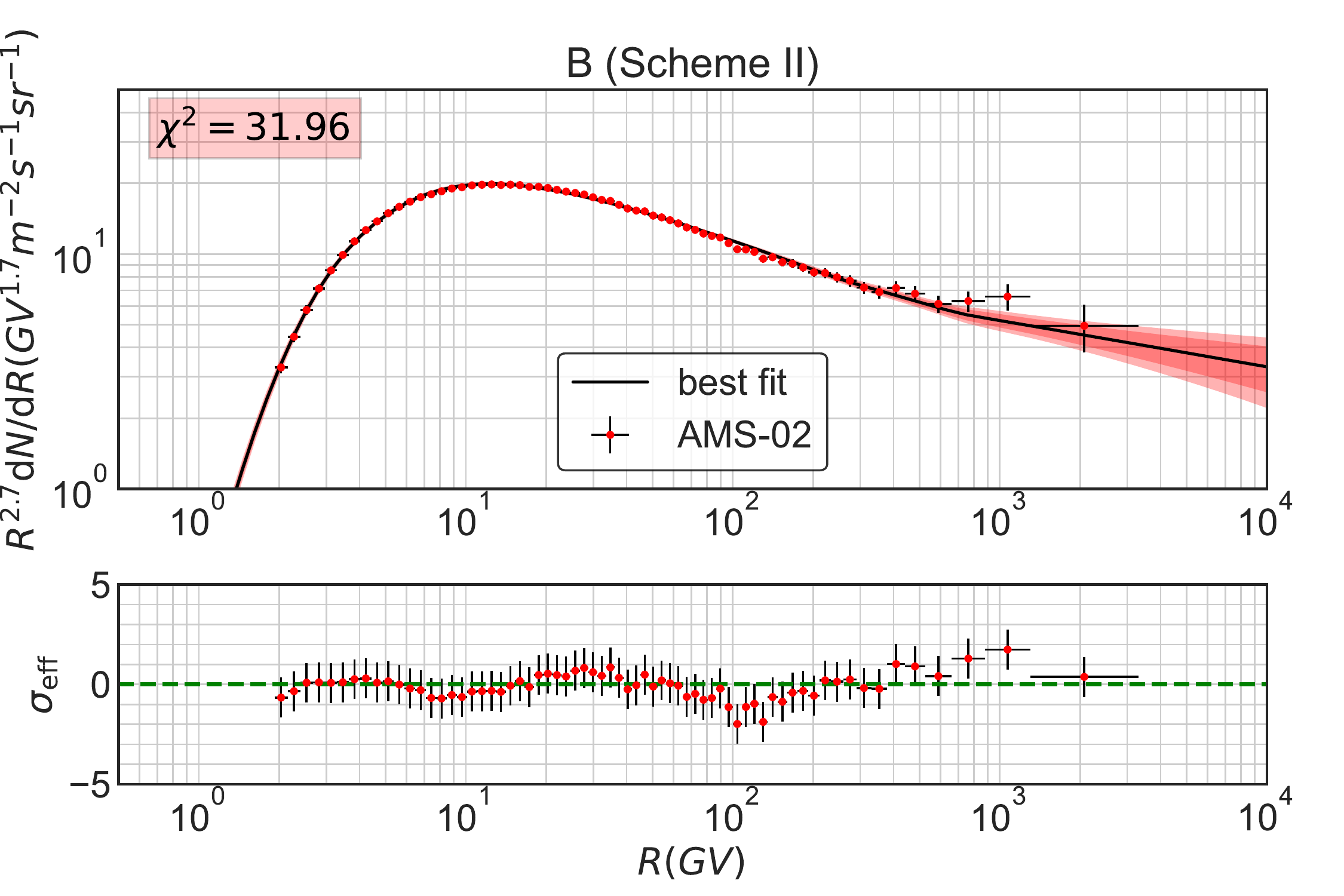}
  \includegraphics[width=0.32\textwidth]{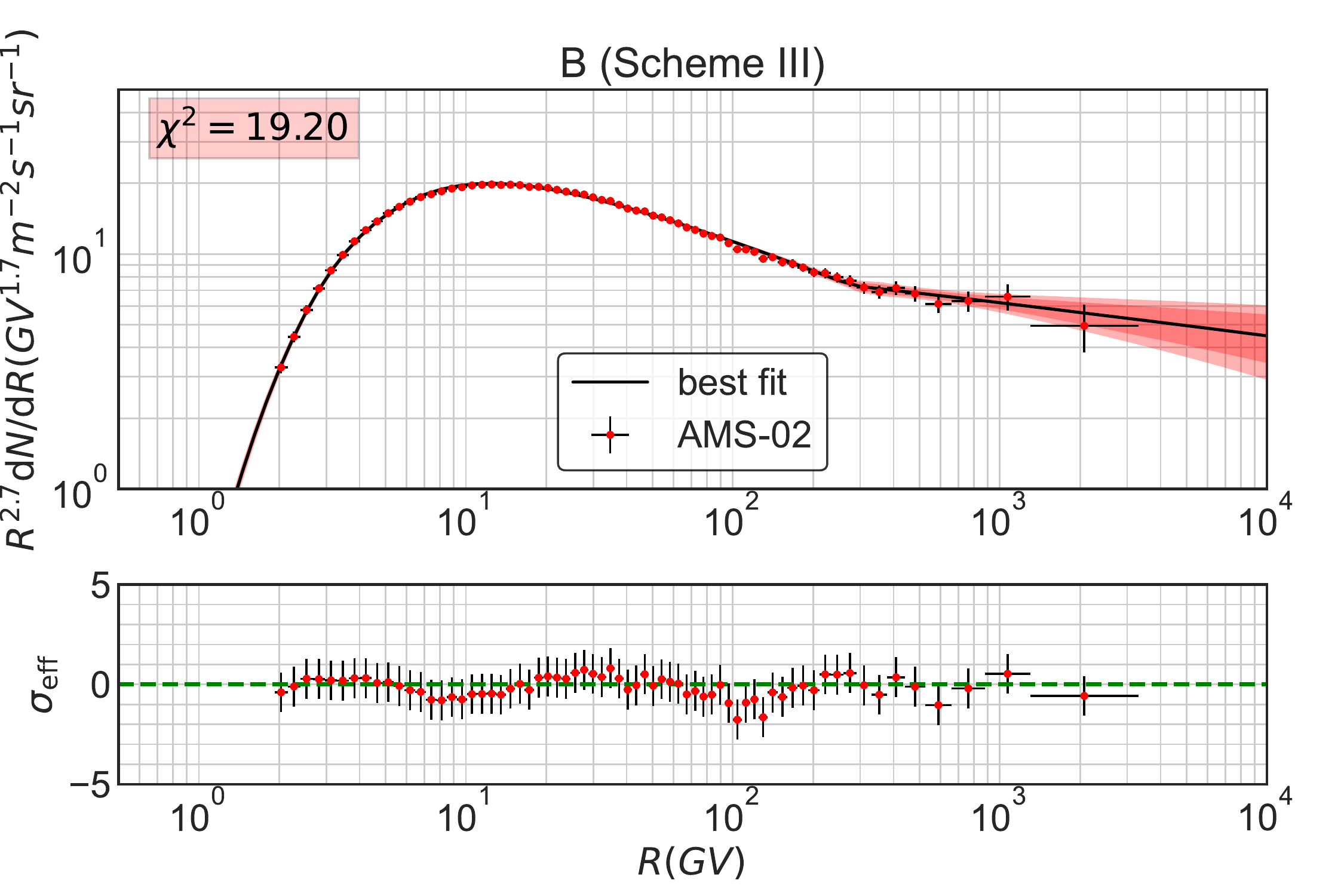}
  \caption{Fitting results and corresponding residuals to the CR spectra of B for the Scheme I, II, and III. The 2$\sigma$ (deep red) and 3$\sigma$ (light red) bounds are also shown in the subfigures. The relevant $\chi^2$ of each spectrum is given in the subfigures as well.}
  \label{fig:spectra03}
\end{figure*}

The samples of the parameters are taken as their posterior probability distribution function (PDF) after the Markov Chains have reached their equilibrium states.\footnote{Here, different prior values are tested to ensure the robustness of the PDFs.} The best-fit results and the corresponding residuals of the spectra are given in Figure \ref{fig:spectra01} (He, C, N, and O), \ref{fig:spectra02} (Ne, Mg, and Si), and \ref{fig:spectra03} (B). 
The best-fit values, statistical mean values and standard deviations, and the 90\% confidence intervals of the parameters in three schemes are shown in Table \ref{tab:params_fitting}. The fitting 1D probability and 2D credible regions (covariances) of posterior PDFs on the parameters of different schemes and groups are collected in Appendix \ref{app1}, \ref{app2}, and \ref{app3}.

\begin{table*}[htbp]
  \caption{Fitting results of the parameters in $\boldsymbol{\theta}_{\mathrm{I}}$, $\boldsymbol{\theta}_{\mathrm{II}}$, and $\boldsymbol{\theta}_{\mathrm{III}}$. Prior: prior interval; Mean/Std: statistical mean and standard deviation values; 90\%: 90\% confidence intervals; Best: best-fit values. }
\scalebox{0.75}[0.75]{
\begin{tabular}{c|c|ccc|ccc|ccc}
  \hline\hline
  \multicolumn{2}{c}{} \vline& \multicolumn{3}{c}{Scheme I} \vline & \multicolumn{3}{c}{Scheme II}  \vline& \multicolumn{3}{c}{Scheme III}  \\
  \hline
ID  &Prior& Mean/Std  & 90\% & Best &Mean/Std  & 90\% & Best & Mean/Std & 90\% & Best\\
  \hline
  $D_{0}\ (10^{28}\cm^{2}\s^{-1})$ & [1, 20]     & 6.6$\pm$0.4   & [5.8, 7.2]   & 6.6  & 5.7$\pm$0.4   & [4.8, 6.5]   & 5.7    & 6.8$\pm$0.6   & [5.7, 7.7]   & 6.9\\
  $\Rbr\ (\GV)$                    & [100, 1000] & 225$\pm$38    & [167, 272]   & 204   & ---           & ---          & ---     & 267$\pm$26    & [226, 312]   & 269\\
  $\delta_{1}$                     & [0.1, 1.0]  & 0.45$\pm$0.01 & [0.43, 0.46] & 0.45  & 0.43$\pm$0.01 & [0.41, 0.45] & 0.43    & 0.44$\pm$0.01 & [0.42, 0.45] & 0.44\\
  $\delta_{2}$                     & [0.1, 1.0]  & 0.31$\pm$0.03 & [0.27, 0.36] & 0.32  & ---           & ---          & ---     & 0.26$\pm$0.02 & [0.22, 0.29] & 0.26\\
  $\eta$                           & [-5.0, 5.0] & -1.5$\pm$0.1  & [-1.8, -1.3] & -1.5 & -1.5$\pm$0.2  & [-1.7, -1.2] & -1.4   & -1.5$\pm$0.1  & [-1.7, -1.4] & -1.5\\
  $z_h\ (\kpc)$                    & [0.5, 20.0] & 10$\pm$1      & [8, 13]      & 10.5  & 7$\pm$1       & [6, 9]       & 7.0     & 11$\pm$2      & [8, 14]      & 10.9\\
  $v_{A}\ (\km/\s)$                & [0, 70]     & 19$\pm$1      & [16, 21]     & 19    & 20$\pm$2      & [18, 23]     & 21      & 20$\pm$1      & [18, 22]     & 20\\
  $\phi\ (\GV)$                    & [0, 1.5]    & 0.72$\pm$0.03 & [0.67, 0.78] & 0.72  & 0.72$\pm$0.03 & [0.68, 0.79] & 0.73    & 0.75$\pm$0.03 & [0.69, 0.81] & 0.75\\
\hline
  $R_{1}^{\mathrm{He}}\ (\GV)$     & [1, 100]    & 4.4$\pm$0.6   & [3.6, 5.6]   & 4.2   & 5.5$\pm$1.2   & [4.1, 6.7]   & 4.8     & 3.6$\pm$0.5   & [2.9, 4.6]   & 3.5\\
  $R_{2}^{\mathrm{He}}\ (\GV)$     & [100, 1000] & 593$\pm$166   & [349, 946]   & 623   & 272$\pm$41    & [220, 371]   & 284     & ---           & ---          &  ---\\
  $\nu_{1}^{\mathrm{He}}$          & [1.0, 4.0]  & 2.78$\pm$0.13 & [2.60, 3.03] & 2.81  & 2.63$\pm$0.10 & [2.49, 2.84] & 2.69    & 3.26$\pm$0.31 & [2.81, 3.84] & 3.24\\
  $\nu_{2}^{\mathrm{He}}$          & [1.0, 4.0]  & 2.34$\pm$0.01 & [2.32, 2.36] & 2.34  & 2.35$\pm$0.01 & [2.33, 2.36] & 2.35    & 2.34$\pm$0.01 & [2.33, 2.35] & 2.34\\
  $\nu_{3}^{\mathrm{He}}$          & [1.0, 4.0]  & 2.25$\pm$0.08 & [2.11, 2.34] & 2.22  & 2.19$\pm$0.03 & [2.13, 2.22] & 2.18    & ---           & ---          &  ---\\
  $R_{1}^{\mathrm{C}}\ (\GV)$      & [1, 100]    & 9$\pm$4       & [3, 17]      & 8     & 7$\pm$5       & [1, 17]      & 5       & 7$\pm$2       & [4, 12]      & 7\\
  $R_{2}^{\mathrm{C}}\ (\GV)$      & [100, 1000] & 455$\pm$118   & [269, 728]   & 448   & 239$\pm$36    & [186, 311]   & 232     & ---           & ---          &  ---\\
  $\nu_{1}^{\mathrm{C}}$           & [1.0, 4.0]  & 2.42$\pm$0.05 & [2.36, 2.54] & 2.43  & 2.45$\pm$0.10 & [2.28, 2.69] & 2.46    & 2.50$\pm$0.08 & [2.40, 2.63] & 2.48\\
  $\nu_{2}^{\mathrm{C}}$           & [1.0, 4.0]  & 2.36$\pm$0.01 & [2.34, 2.37] & 2.36  & 2.37$\pm$0.01 & [2.35, 2.39] & 2.37    & 2.36$\pm$0.01 & [2.34, 2.37] & 2.36\\
  $\nu_{3}^{\mathrm{C}}$           & [1.0, 4.0]  & 2.24$\pm$0.07 & [2.10, 2.32] & 2.24  & 2.18$\pm$0.04 & [2.12, 2.23] & 2.18    & ---           & ---          &  ---\\
  $R_{1}^{\mathrm{N}}\ (\GV)$      & [1, 100]    & 70$\pm$16     & [37, 98]     & 75    & 16$\pm$9      & [3, 38]      & 15      & 78$\pm$21     & [29, 99]     & 84\\
  $R_{2}^{\mathrm{N}}\ (\GV)$      & [100, 1000] & 822$\pm$125   & [552, 981]   & 817   & 164$\pm$28    & [118, 209]   & 160     & ---           & ---          &  ---\\
  $\nu_{1}^{\mathrm{N}}$           & [1.0, 4.0]  & 2.40$\pm$0.03 & [2.35, 2.44] & 2.40  & 2.34$\pm$0.10 & [2.16, 2.51] & 2.37    & 2.40$\pm$0.03 & [2.35, 2.45] & 2.40\\
  $\nu_{2}^{\mathrm{N}}$           & [1.0, 4.0]  & 2.28$\pm$0.04 & [2.22, 2.37] & 2.29  & 2.44$\pm$0.04 & [2.40, 2.52] & 2.44    & 2.29$\pm$0.04 & [2.21, 2.35] & 2.29\\
  $\nu_{3}^{\mathrm{N}}$           & [1.0, 4.0]  & 1.86$\pm$0.33 & [1.32, 2.22] & 1.71  & 2.00$\pm$0.06 & [1.90, 2.10] & 2.00    & ---           & ---          &  ---\\
  $R_{1}^{\mathrm{O}}\ (\GV)$      & [1, 100]    & 6$\pm$3       & [2, 11]      & 5     & 8$\pm$3       & [2, 15]      & 7       & 5$\pm$2       & [3, 8]       &  5\\
  $R_{2}^{\mathrm{O}}\ (\GV)$      & [100, 1000] & 767$\pm$125   & [504, 961]   & 759   & 696$\pm$117   & [431, 871]   & 642     & ---           & ---          &  ---\\
  $\nu_{1}^{\mathrm{O}}$           & [1.0, 4.0]  & 2.47$\pm$0.15 & [2.21, 2.75] & 2.46  & 2.47$\pm$0.07 & [2.37, 2.65] & 2.47    & 2.58$\pm$0.12 & [2.42, 2.83] & 2.55\\
  $\nu_{2}^{\mathrm{O}}$           & [1.0, 4.0]  & 2.38$\pm$0.01 & [2.37, 2.40] & 2.38  & 2.38$\pm$0.01 & [2.36, 2.40] & 2.38    & 2.38$\pm$0.01 & [2.37, 2.40] & 2.38\\
  $\nu_{3}^{\mathrm{O}}$           & [1.0, 4.0]  & 2.20$\pm$0.12 & [1.99, 2.35] & 2.17  & 2.02$\pm$0.09 & [1.85, 2.16] & 2.02    & ---           & ---          &  ---\\
  \hline
  $R_{1}^{\mathrm{Ne}}\ (\GV)$     & [1, 100]    & 8$\pm$2       & [6, 12]      & 8     & 9$\pm$3       & [6, 14]      & 9       & 12$\pm$5      & [7, 21]      & 10\\
  $R_{2}^{\mathrm{Ne}}\ (\GV)$     & [100, 1000] & 797$\pm$119   & [566, 980]   & 823   & 788$\pm$88    & [586, 941]   & 764     & ---           & ---          &  ---\\
  $\nu_{1}^{\mathrm{Ne}}$          & [1.0, 4.0]  & 2.11$\pm$0.09 & [1.92, 2.26] & 2.13  & 2.18$\pm$0.08 & [2.01, 2.30] & 2.18    & 2.21$\pm$0.08 & [2.07, 2.33] & 2.23\\
  $\nu_{2}^{\mathrm{Ne}}$          & [1.0, 4.0]  & 2.38$\pm$0.01 & [2.36, 2.40] & 2.38  & 2.38$\pm$0.01 & [2.36, 2.40] & 2.38    & 2.38$\pm$0.01 & [2.37, 2.40] & 2.38\\
  $\nu_{3}^{\mathrm{Ne}}$          & [1.0, 4.0]  & 2.15$\pm$0.21 & [1.72, 2.41] & 2.08  & 1.87$\pm$0.15 & [1.54, 2.08] & 1.82    & ---           & ---          &  ---\\
  $R_{1}^{\mathrm{Mg}}\ (\GV)$     & [1, 100]    & 28$\pm$11     & [9, 48]      & 27    & 14$\pm$7      & [3, 29]      & 11      & 44$\pm$25     & [7, 89]      & 40\\
  $R_{2}^{\mathrm{Mg}}\ (\GV)$     & [100, 1000] & 550$\pm$186   & [217, 939]   & 559   & 385$\pm$85    & [240, 592]   & 398     & ---           & ---          &  ---\\
  $\nu_{1}^{\mathrm{Mg}}$          & [1.0, 4.0]  & 2.41$\pm$0.03 & [2.37, 2.45] & 2.41  & 2.38$\pm$0.05 & [2.25, 2.46] & 2.38    & 2.43$\pm$0.02 & [2.38, 2.46] & 2.43\\
  $\nu_{2}^{\mathrm{Mg}}$          & [1.0, 4.0]  & 2.46$\pm$0.02 & [2.44, 2.49] & 2.46  & 2.46$\pm$0.01 & [2.44, 2.48] & 2.46    & 2.47$\pm$0.02 & [2.44, 2.50] & 2.47\\
  $\nu_{3}^{\mathrm{Mg}}$          & [1.0, 4.0]  & 2.40$\pm$0.14 & [2.10, 2.63] & 2.39  & 2.30$\pm$0.11 & [2.09, 2.44] & 2.27    & ---           & ---          &  ---\\
  $R_{1}^{\mathrm{Si}}\ (\GV)$     & [1, 100]    & 42$\pm$13     & [22, 70]     & 42    & 53$\pm$10     & [34, 74]     & 57      & 55$\pm$20     & [22, 89]     & 53\\
  $R_{2}^{\mathrm{Si}}\ (\GV)$     & [100, 1000] & 528$\pm$156   & [236, 858]   & 534   & 355$\pm$116   & [168, 641]   & 315     & ---           & ---          &  ---\\
  $\nu_{1}^{\mathrm{Si}}$          & [1.0, 4.0]  & 2.38$\pm$0.02 & [2.34, 2.41] & 2.38  & 2.39$\pm$0.02 & [2.36, 2.41] & 2.39    & 2.39$\pm$0.02 & [2.35, 2.42] & 2.39\\
  $\nu_{2}^{\mathrm{Si}}$          & [1.0, 4.0]  & 2.45$\pm$0.02 & [2.43, 2.48] & 2.45  & 2.46$\pm$0.02 & [2.44, 2.52] & 2.47    & 2.46$\pm$0.02 & [2.43, 2.49] & 2.46\\
  $\nu_{3}^{\mathrm{Si}}$          & [1.0, 4.0]  & 2.48$\pm$0.11 & [2.30, 2.68] & 2.47  & 2.32$\pm$0.08 & [2.10, 2.41] & 2.31    & ---           & ---          &  ---\\
  \hline
  $N_{\mathrm{He}}/71990.0$        & [0.1,5.0]   & 1.40$\pm$0.01 & [1.38, 1.42] & 1.40  & 1.39$\pm$0.01 & [1.38, 1.42] & 1.40    & 1.40$\pm$0.01 & [1.39, 1.42] & 1.40\\
  $N_{\mathrm{C}}/2819.0$          & [0.1,5.0]   & 1.21$\pm$0.01 & [1.19, 1.23] & 1.21  & 1.20$\pm$0.02 & [1.18, 1.22] & 1.19    & 1.22$\pm$0.01 & [1.20, 1.24] & 1.22\\
  $N_{\mathrm{N}}/182.8$           & [0.1,5.0]   & 1.45$\pm$0.05 & [1.37, 1.51] & 1.44  & 1.35$\pm$0.07 & [1.25, 1.46] & 1.36    & 1.45$\pm$0.04 & [1.37, 1.51] & 1.45\\
  $N_{\mathrm{O}}/3822.0$          & [0.1,5.0]   & 1.11$\pm$0.01 & [1.09, 1.13] & 1.11  & 1.12$\pm$0.01 & [1.10, 1.14] & 1.12    & 1.11$\pm$0.01 & [1.10, 1.13] & 1.11\\
  $N_{\mathrm{Ne}}/312.5$          & [0.1,5.0]   & 1.59$\pm$0.03 & [1.54, 1.63] & 1.59  & 1.62$\pm$0.03 & [1.57, 1.66] & 1.62    & 1.60$\pm$0.02 & [1.56, 1.64] & 1.60\\
  $N_{\mathrm{Mg}}/658.1$          & [0.1,5.0]   & 0.94$\pm$0.02 & [0.91, 0.96] & 0.93  & 0.95$\pm$0.02 & [0.92, 0.98] & 0.95    & 0.94$\pm$0.02 & [0.91, 0.96] & 0.94\\
  $N_{\mathrm{Si}}/725.7$          & [0.1,5.0]   & 1.02$\pm$0.02 & [0.99, 1.05] & 1.02  & 1.02$\pm$0.02 & [0.98, 1.05] & 1.02    & 1.01$\pm$0.01 & [1.00, 1.04] & 1.02\\
 \hline
  \multicolumn{2}{c}{$\chi^{2}/\text{d.o.f.}$} \vline& \multicolumn{3}{c}{78.6/484}  \vline& \multicolumn{3}{c}{118.0/486}  \vline& \multicolumn{3}{c}{105.2/498}  \\
\hline\hline
\end{tabular}
}
\label{tab:params_fitting}
\end{table*}

For Scheme I, II, and III, we have $\chi^{2}_{\mathrm{I}}/d.o.f = 78.6/484 \approx 0.16$, $\chi^{2}_{\mathrm{II}}/d.o.f = 118.0/486 \approx 0.24$, and $\chi^{2}_{\mathrm{III}}/d.o.f = 105.2/498 \approx 0.21$ for the best-fit results, respectively.

  In Bayesian terms, the criterion of a decisive evidence between 2 models is $\Delta \chi^{2} \geq 10$ (see, e.g., \citet{Genolini2017}), with the same $d.o.f$.
  Comparing Scheme II and III, $\Delta \chi^{2} = \chi^{2}_{\mathrm{II}} - \chi^{2}_{\mathrm{III}} = 12.8$, and the $d.o.f$ of Scheme II is even smaller than that of Scheme III simultaneously, which is a decisive evidence and indicates that the Scheme III is statistically significant better than Scheme II in current data set.
  It is consistent with some previous works (see, e.g., \citet{Genolini2017,Niu2020}), which declare that the AMS-02 nuclei data favor the spectral hardening coming from the propagation process rather than the CR primary source.
  Comparing Scheme I and III, although the $d.o.f$ of Scheme I is smaller than that of Scheme III (caused by additional 14 parameters), the $\Delta \chi^{2} = \chi^{2}_{\mathrm{III}} - \chi^{2}_{\mathrm{I}} = 26.6$ is really a large improvement. On the other hand, considering the differences of $\chi^{2}/d.o.f$, because that between Scheme II and III $\sim 0.03$ is statistically significant, that between Scheme I and III $\sim 0.05$ indicates that the Scheme I is statistically significant better than Scheme III in current data set.

The small values of $\chi^{2}/d.o.f$ in the three schemes are mainly caused by the correlations of the systematic errors of the data. More appropriate treatment of the systematic errors can be found in \citet{Derome2019,Weinrich202001,Heisig2020,Korsmeier2021}.

In Figure \ref{fig:spectra01}, \ref{fig:spectra02}, and \ref{fig:spectra03}, for a specific nuclei spectrum,
  Scheme I gives the smallest $\chi^{2}$ in most cases (which is due to its precise description of the high-rigidity spectral structures), while Scheme II and III give larger $\chi^{2}$, because both of them can not precisely reproduce the spectral breaks around 200 GV and 400-800 GV simultaneously.
  Comparing the $\chi^{2}$ of Scheme II and III for different species in Figure \ref{fig:spectra01}, Scheme II gives out larger $\chi^{2}$ in the case of He and O, and Scheme III gives out larger $\chi^{2}$ in the case of C and N, which indicates that the spectral breaks around 200 GV and 400-800 GV have different weights for different nuclei species. 
  Comparing the $\chi^{2}$ of Scheme II and III for different species in Figure \ref{fig:spectra02}, Scheme II always gives out larger $\chi^{2}$, it indicates that the spectral breaks around 400-800 GV are not that important in the case of Ne, Mg, and Si, which represents that Ne, Mg, and Si and He, C, and O might be two different classes of primary CRs \citep{AMS02_Ne_Mg_Si}.
  Comparing the $\chi^{2}$ of Scheme II and III for B in Figure \ref{fig:spectra03}, Scheme II gives out $\chi^{2}=31.96$ and Scheme III gives out $\chi^{2}=19.20$ ($\Delta \chi^{2} \sim 13$), which indicates that the spectral break around 200 GV of B is its dominating feature and it favors the propagation origin of the spectral hardening.

  The detailed information about the three schemes can be read out in Table \ref{tab:params_fitting}.
  The propagation parameters in Scheme I and III have similar distributions, except the cases of $R_{br}$ and $\delta_2$. Because that in Scheme I is only responsible for the spectral breaks around 200 GV, while that in Scheme III needs to reproduce the spectral breaks around 200 GV and 400-800 GV simultaneously. The distributions of $D_{0}$ and $z_{h}$ are slightly different in Scheme II and Scheme I/III. This is because these two parameters are mainly determined by the spectrum of B, which is hardening around 200 GV more than the primary species. In Scheme I/III, the spectrum of B is precisely reproduced with the diffusion break and $\delta_{1}$ and $\delta_{2}$, while that is roughly reproduced without the diffusion break in Scheme II, and it influences the distributions of $D_{0}$ and $z_{h}$.
  The solar-modulation potential $\phi$ in this work has a range from 0.67 GV to 0.81 GV, which is a bit larger than $\phi = 0.64 \GV$ based the the NEWK\footnote{\url{http://www01.nmdb.eu/station/newk/}} neutron monitor experiment from Cosmic-Ray DataBase (CRDB\footnote{\url{https://lpsc.in2p3.fr/crdb/}}) \citep{Ghelfi2016,Ghelfi2017}. Considering that it is an effective value which is coupled with $v_{A}$ and $\eta$ and does not impact on the discussion regarding the high-rigidity breaks, we will not discuss this issue in depth in this work.

  About the spectral parameters, comparing the high-rigidity break positions in injection spectra with and without the diffusion break (i.e. $R_{2}$ in Scheme I and II, respectively), we find that with the diffusion break (in Scheme I) always have larger values, in which case the high-rigidity breaks just take charge of the spectral breaks about 400-800 GV, while both the spectral breaks around 200 GV and 400-800 GV determines the high-rigidity breaks without the diffusion break in Scheme II. For the high-rigidity break positions in diffusion coefficients with and without additional high-rigidity breaks in injection spectra (i.e. $\Rbr$ in Scheme I and III, respectively), that with the additional high-rigidity breaks in injection spectra (in Scheme I) has a smaller value, which accounts for the spectral breaks around 200 GV. However, that in Scheme III accounts for both the spectral breaks around 200 GV and 400-800 GV, then has a larger value. For the differences between spectral index in the injection spectra above and below the high-rigidity breaks (i.e. $\Delta \nu \equiv \vert \nu_{3} - \nu_{2}  \vert$ in Scheme I and II), that in Scheme II have larger values in most cases, which represents that the hardening in most of the spectra around 200 GV and 400-800 GV is taken up by $\Delta \nu$ alone in Scheme II, while it is shared by $\Delta \nu$ and the break in diffusion coefficient simultaneously in Scheme I. The exception comes from the spectra of N, in which case the $\Delta \nu$ in Scheme I has larger value. It comes from the sudden hardening of its spectra around 800 GV, which cannot be precisely reproduced in Scheme II.

Hereafter, we focus on the fitting results of Scheme I.

\section{Discussions and Conclusion}

\begin{figure*}[htp!]
  \gridline{\fig{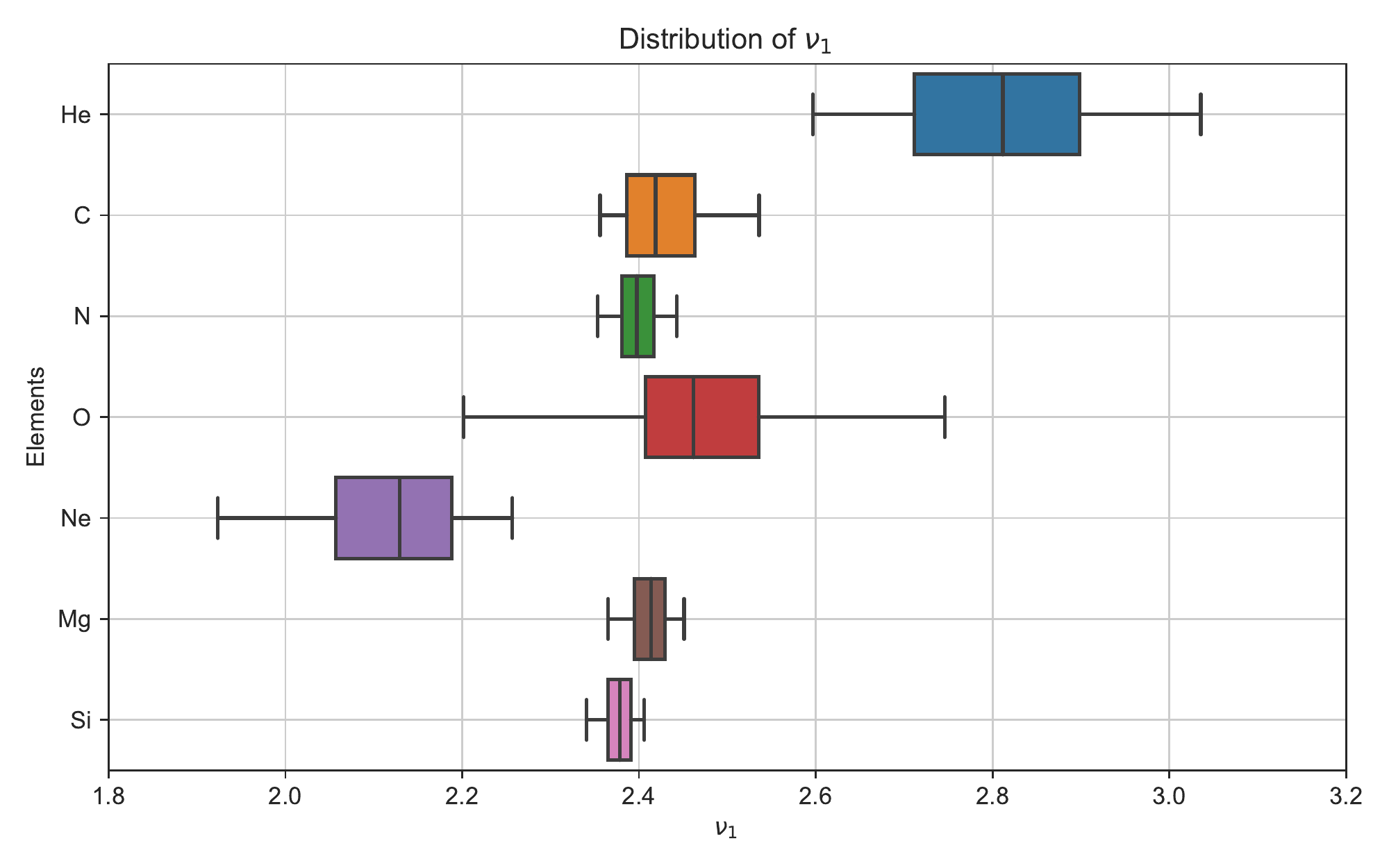}{0.33\textwidth}{(a)}
    \fig{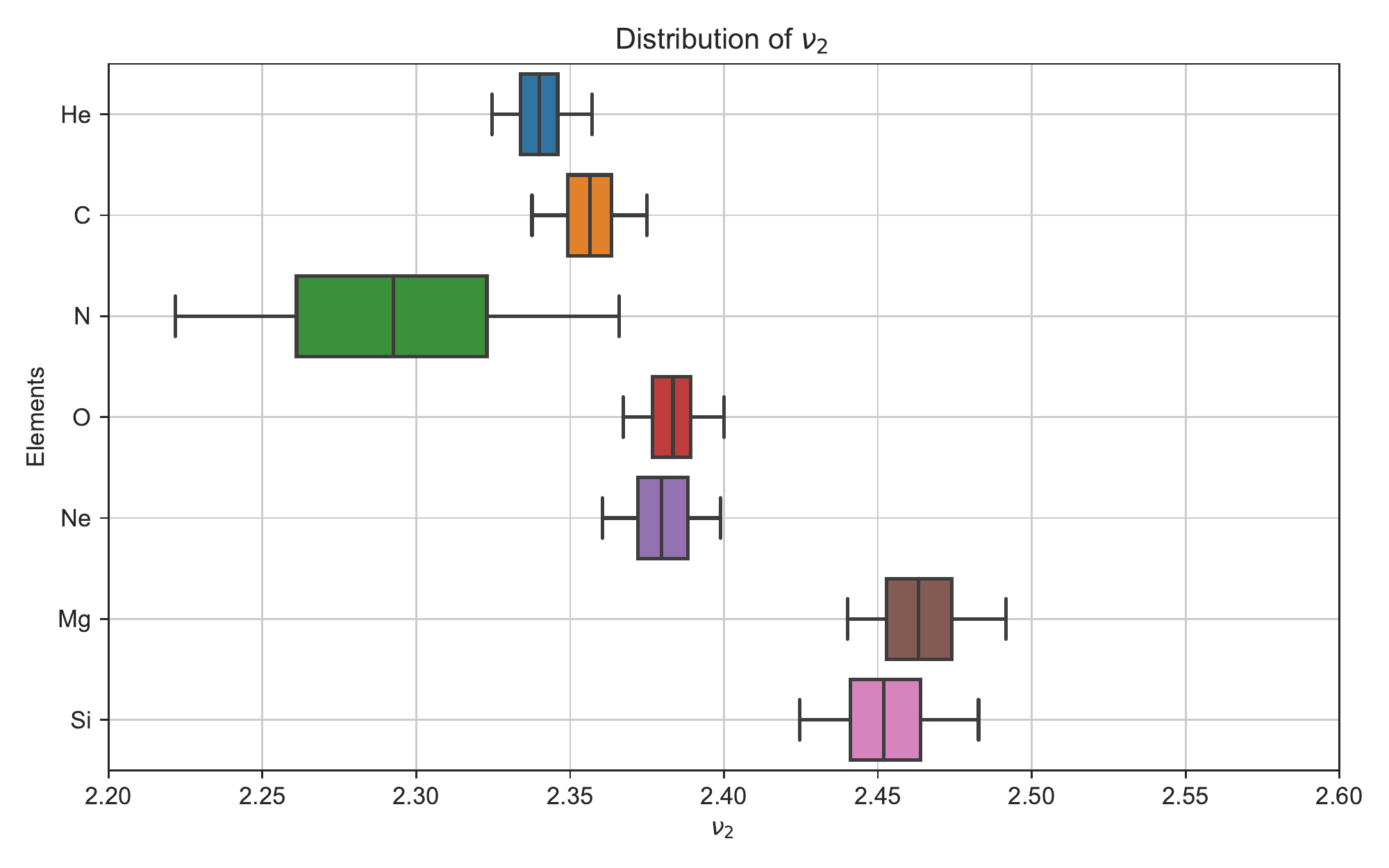}{0.33\textwidth}{(b)}
    \fig{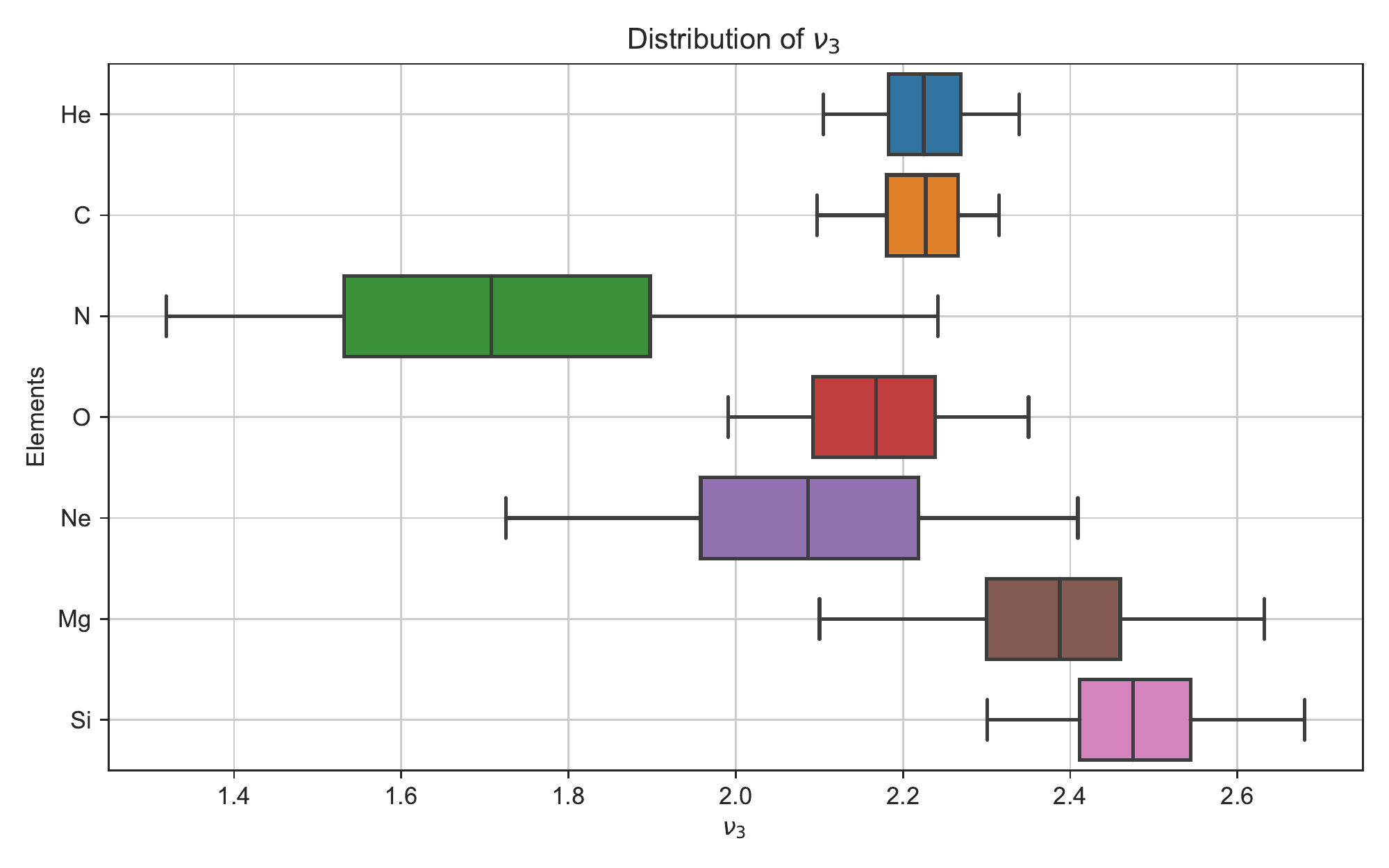}{0.33\textwidth}{(c)}}
  \gridline{\fig{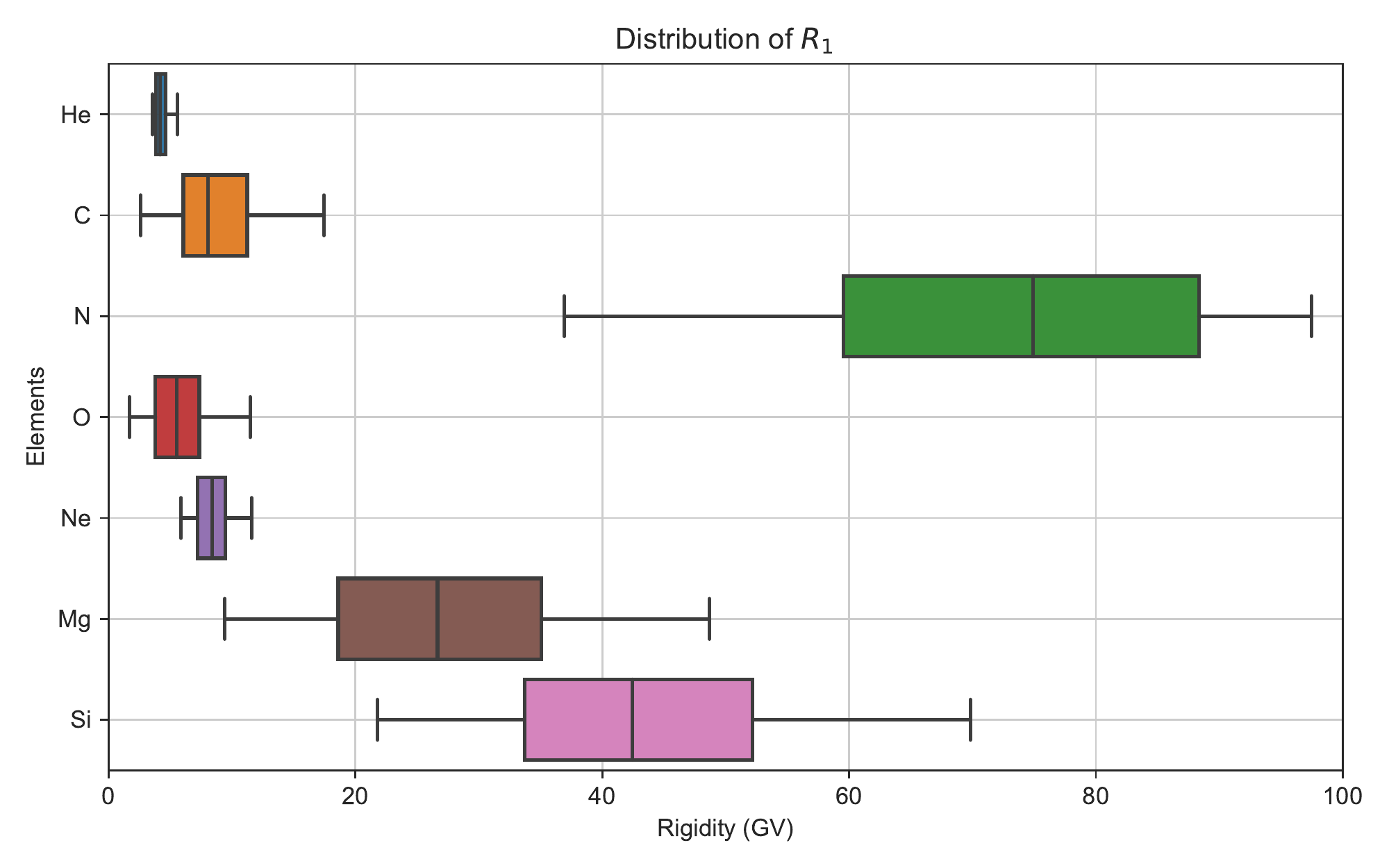}{0.33\textwidth}{(d)}
    \fig{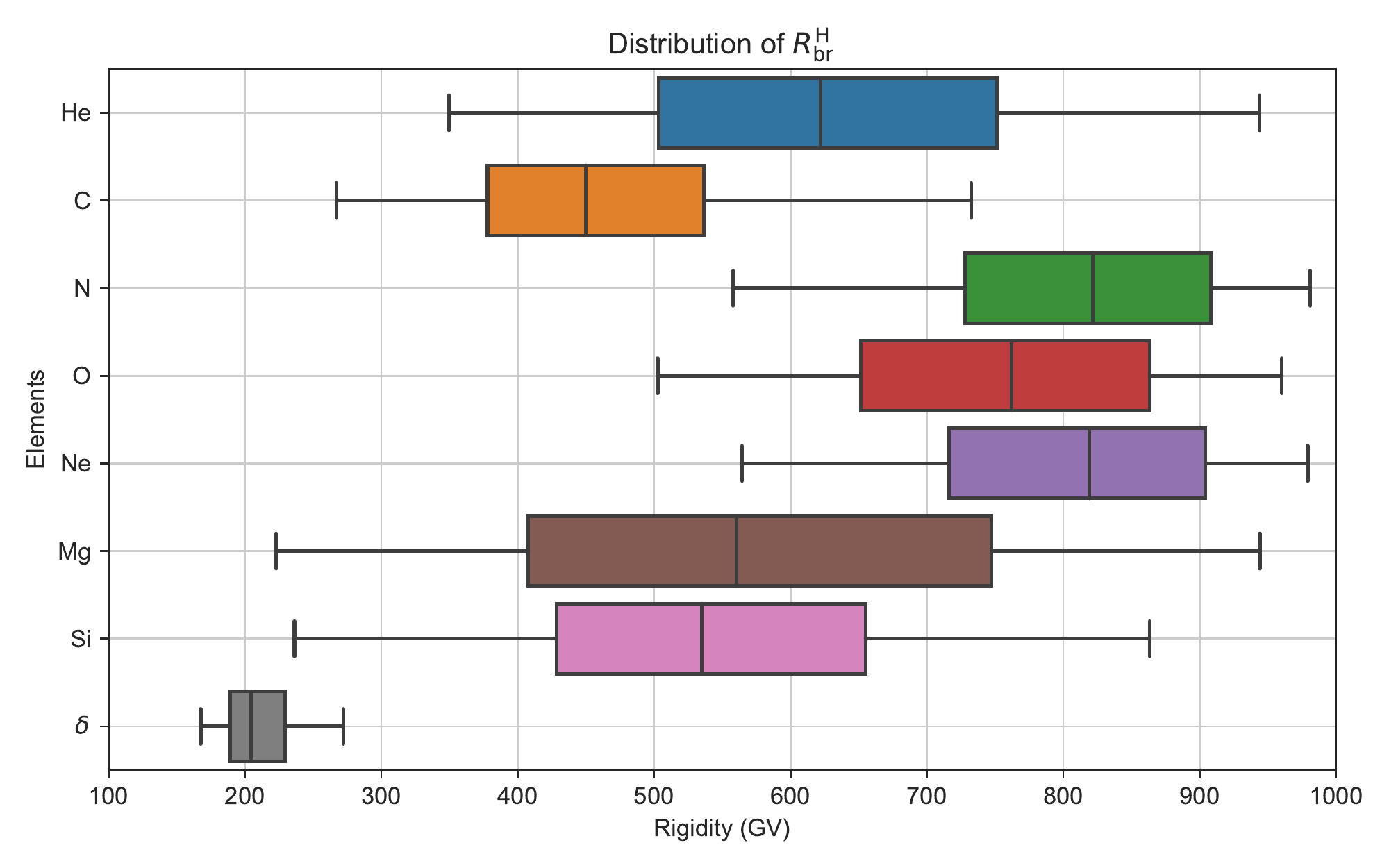}{0.33\textwidth}{(e)}}
  \caption{Boxplots for the primary source injection parameters of different species in Scheme I. The band inside the box shows the median value of the dataset, the box shows the quartiles, and the whiskers extend to show the rest of the distribution which are edged by the 5th percentile and the 95th percentile.}
  \label{fig:spec_params}
\end{figure*}

In order to compare the primary source injection parameters of different species, the box plot of these parameters ($\nu_{1}$, $\nu_{2}$, $\nu_{3}$, $R_{1}$, and $R_{2}$) are shown in Figure \ref{fig:spec_params}.

The large deviations and uncertainties of N compared with other species in subfigures (b) and (c) are related to its hybrid origins (which is expected to contain both primary and secondary components), while the production cross sections of its secondary components are not precisely provided in {\sc galprop} v56. Unless specifically mentioned, the following discussions exclude the fitting results of N.

In subfigure (a), the $\nu_{1}$ of He and Ne show large deviations compared with other species; in subfigure (b), the distributions of $\nu_{2}$ indicate that He and C should be in a group, O and Ne should be in a group, and Mg and Si should be in a group; in subfigure (c), the values of  $\nu_{3}$ of Mg and Si have some deviations compared with other species; in subfigure (d), the values of  $R_{1}$ of Mg and Si also show some deviations compared with other species; in subfigure (e), it shows that O and Ne have values of $R_{2}$ (Here, $R_{2} \equiv R_{\mathrm{br}}^{\mathrm{H}}$ for He, C, N, O, Ne, Mg, and Si) with large overlaps compared with that of He, C, Mg, and Si. Taken together, the CR species Mg and Si might have similar origins because of their similar distributions of primary source injection parameters.
Another hint should be noted is that the relationships between different species could be different in low and high-rigidity regions. For example, He and Ne, show similar $\nu_{2}$, $\nu_{3}$, $R_{2}$ distributions to C and O respectively in high-rigidity region, but show large $\nu_{1}$ deviations to C and O respectively in low-rigidity region. This might be some hints that the CR species He and Ne have different origins in low-rigidity regions.

\begin{figure*}[!htbp]
   \centering
   \includegraphics[width=0.49\textwidth]{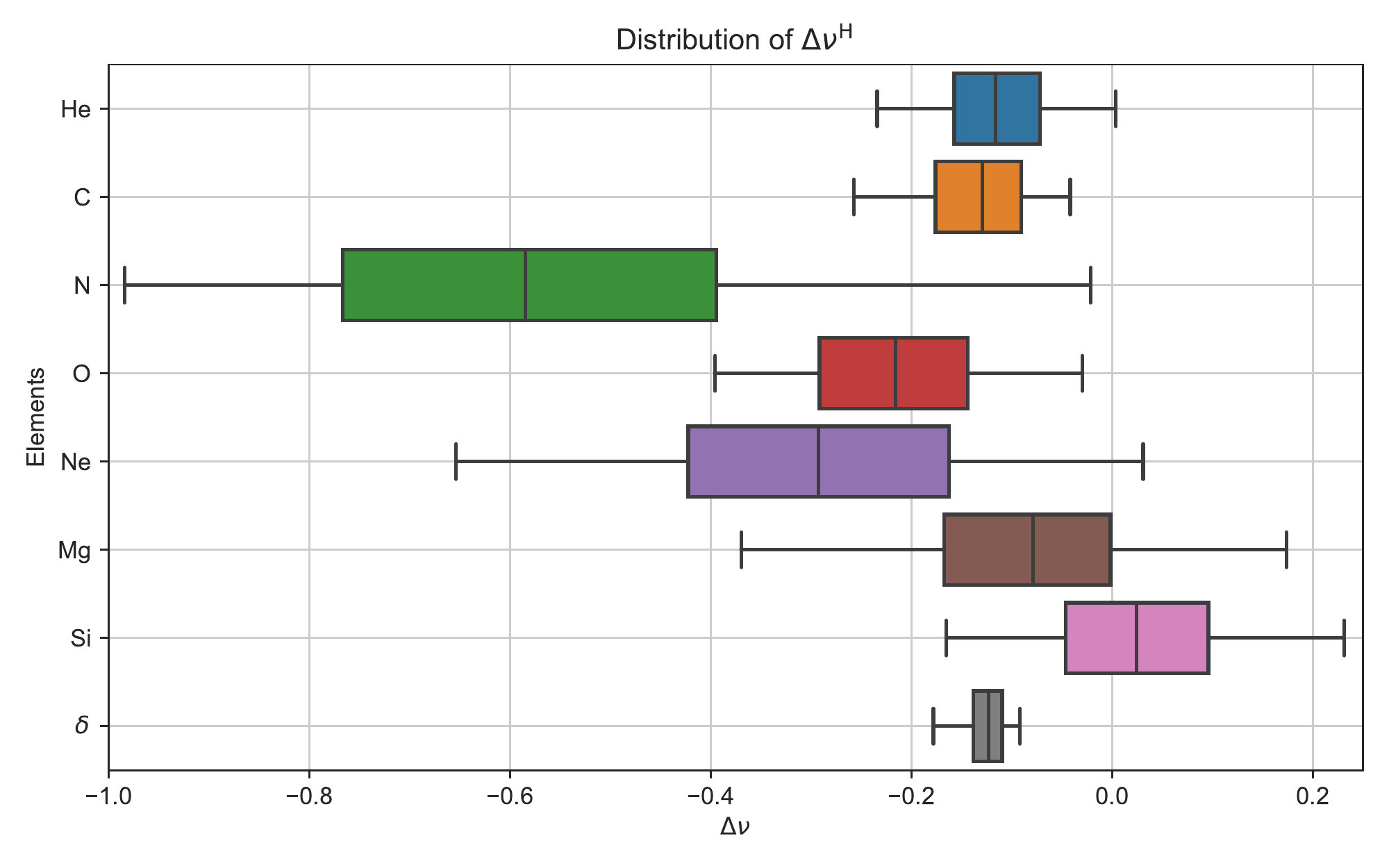}
   \caption{Boxplots for $\Delta \nu^{\mathrm{H}}$. The band inside the box shows the median value of the dataset, the box shows the quartiles, and the whiskers extend to show the rest of the distribution which are edged by the 5th percentile and the 95th percentile.}
   \label{fig:breaks_nu_high}
 \end{figure*}

 In order to explore the properties of the spectral hardening in 100-1000 GV, the posterior mean and standard deviation of the high-rigidity break ($R_{\mathrm{br}}^{\mathrm{H}} \equiv R_{2}^i$ for He, C, N, O, Ne, Mg, and Si; $R_{\mathrm{br}}^{\mathrm{H}} \equiv R_{\mathrm{br}}$ for $\delta$) and the differences between the spectral index above and below it ($\Delta \nu^{\mathrm{H}} \equiv \nu_{3}^{i} - \nu_{2}^{i}$ for He, C, N, O, Ne, Mg, and Si; $\Delta \nu^{\mathrm{H}} \equiv \delta_2 - \delta_1 $ for $\delta$) are summarized in Table \ref{tab:spectra_params}. The box plot of these two kinds of parameters are shown in subfigure (e) of Figure \ref{fig:spec_params} and Figure \ref{fig:breaks_nu_high}, respectively.

 \begin{table}
   \caption{Posterior mean and standard deviation of $R_{\mathrm{br}}^{\mathrm{H}}$ and $\Delta \nu^{\mathrm{H}}$.}
   \begin{center}
     \begin{tabular}{c|cc}
       \hline\hline
       ID  &$R_{\mathrm{br}}^{\mathrm{H}}$ (GV) & $\Delta \nu^{\mathrm{H}}$  \\
       \hline
       $\delta$ &225 $\pm$ 38 &-0.13 $\pm$ 0.03 \\
       He &593 $\pm$ 166  &-0.12 $\pm$ 0.07 \\
       C &455 $\pm$ 118  &-0.14 $\pm$ 0.07 \\
       N &822 $\pm$ 125 &-0.56 $\pm$ 0.31 \\
       O &767 $\pm$ 125 &-0.22 $\pm$ 0.11 \\
       Ne &797 $\pm$ 119 &-0.30 $\pm$ 0.21 \\
       Mg &550 $\pm$ 186 &-0.08 $\pm$ 0.16 \\
       Si &528 $\pm$ 156 &0.02 $\pm$ 0.13 \\
       \hline\hline
     \end{tabular}
   \end{center}
   \label{tab:spectra_params}
 \end{table}

  In the subfigure (e) of Figure \ref{fig:spec_params}, the high-rigidity breaks show different distributions: for He, C, Mg and Si, $R_{\mathrm{br}}^{\mathrm{H}}$ mainly distributes less than about 700 GV; for N, O and Ne, it almost distributes greater than 700 GV. These different distributions of the high-rigidity breaks cannot be naturally reproduced by a uniform acceleration mechanism in the primary source injection spectra for different CR nuclei species. As some previous work have been pointed out (see, e.g., \citet{Yuan2011,Yue2019,Yuan2020,Niu2021cpc}), it could be naturally explained by the superposition of different kinds of sources. In this scenario, each kind of the sources have similar spectral indices for all the primary source injection but have different element abundances between different kinds of sources.\footnote{An interesting and detailed work on revealing the origin of galactic CRs by their composition has been proposed in \citet{Tatischeff2021}.}  Different from the $R_{\mathrm{br}}^{\mathrm{H}}$ of the primary source injection spectra which have large overlaps between each other, that of the diffusion coefficient demonstrates little uncertainty and has large deviation to others. It indicates the necessity of employing a break in the diffusion coefficient, which is the observational evidence of the propagation origin scenarios (such as in \citet{Blasi2012,Tomassetti2012,Tomassetti2015apjl01,Tomassetti2015prd,Feng2016,Genolini2017,Jin2016CPC,Guo2018cpc,Guo2018prd,Liu2018,Niu2019,Boschini2020apj,Boschini2020apjs}).

 In Figure \ref{fig:breaks_nu_high}, except the quite large uncertainty for N (which is caused by its primary/secondary hybrid origin), for He, C, O, and Ne, $\Delta \nu^{\mathrm{H}}$ has a confidence level of about 95\% smaller than 0, which are the signs of the necessity of hardening contributions from the primary source injection at about 400-800 GV. For Mg and Si, the $\Delta \nu^{\mathrm{H}}$ distributions around 0 (which can also be noted in Table \ref{tab:spectra_params}) indicate that it is not necessary to employ a high-rigidity break to reproduce the spectral hardening at about 400-800 GV for them. This result is also consistent with the above analysis that Mg and Si should be grouped together and their CRs might have similar origins.
On the other hand, the concentrated distribution of $\Delta \nu^{\mathrm{H}}$ for  $\delta$ also shows its necessity to reproduce the data set, whose value of $\sim -0.1$ also has been proved by some of the previous works based on different configurations (see, e.g., \citet{Genolini2017,Genolini2019,Niu2020}).

In summary, if we want to reproduce the spectral hardening in the CR nuclei species at a few hundred GV precisely, not only an extra break at about 200 GV in the diffusion coefficient is needed (see, e.g., \citet{Genolini2017,Genolini2019,Niu2019,Niu2020}), but the extra independent high-rigidity breaks at about 400-800 GV in the primary source injection spectra for different CR species are also needed (see, e.g., \citet{Niu2021cpc,Korsmeier2021a}). The result shows statistically significant improvement compared with the schemes which use a break in the diffusion coefficient or breaks in the primary source injection alone to reproduce the AMS-02 CR nuclei spectra.
The break in the diffusion coefficient could come from the propagation process, which can be reproduced by the spatial-dependent propagation (see, e.g., \citet{Tomassetti2012,Guo2016,Feng2016}). The different propagation regions of the galactic CRs are corresponding to the structures of our galaxy (i.e., the galaxy center, the bulk, the disk, and the halo), which have different densities of ISM and thus different propagation environments.
The different breaks in the primary source injection spectra could come from the superposition of different kinds of sources. On one hand, these different kinds of sources can be corresponding to the galactic averaged CR sources and a local CR source (such as Geminga SNR \citep{Zhao2022}). On the other hand, it also can be correspond to different kinds of CR factories: such as the different population of supernova remnants \citep{Aharonian2004}, galactic center \citep{Scherer2022}, novas \citep{Nova_CR}, etc. In any case, as long as they have different elemental abundances (which is natural), it will produce different breaks and spectral indices. Of course, a combination of the above two situations is also possible (see, e.g., \citet{Zhang2022}).
Consequently, the CR nuclei spectral hardening at a few hundred GV has hybrid origins.

Moreover, in low-rigidity regions, $\nu_{1}$ for He and Ne show large deviations to other nuclei species, which indicates their different CR origins and the CR universality is violated in all the rigidity region from sub-GV to TV. The precise CR spectra data reveals a more complicated CR nuclei origin than we thought, and it will be clearer in the future based on more precise data.


\begin{acknowledgments}
This research was supported by the National Natural Science Foundation of China (NSFC) (No. 12005124 and No. 12147215) and the Applied Basic Research Programs of Natural Science Foundation of Shanxi Province (No. 201901D111043).
\end{acknowledgments}

\software{{\tt emcee} \citep{emcee}, {galprop} \citep{Strong1998,Moskalenko2002,Strong2001,Moskalenko2003,Ptuskin2006}, {corner} \citep{corner}, {seaborn} \citep{seaborn}}

  The data of the posterior samples of the parameter for three schemes is available on Zenodo under an open-source Creative Commons Attribution license: \dataset[doi:10.5281/zenodo.6435163]{https://doi.org/10.5281/zenodo.6435163}.


\appendix
\restartappendixnumbering

The best-fit results and the corresponding residuals of the spectra are given in Appendix Figure \ref{fig:spectra01} (He, C, N, and O), \ref{fig:spectra02} (Ne, Mg, Si), and \ref{fig:spectra03} (B). Note that in the lower panel of subfigures in Fig. \ref{fig:spectra01},  \ref{fig:spectra02}, and  \ref{fig:spectra03}, the $\sigma_{\mathrm{eff}}$ is defined as
\begin{equation}
  \sigma_{\mathrm{eff}} = \frac{f_{\mathrm{obs}} - f_\mathrm{cal}}{\sqrt{\sigma_\mathrm{stat}^{2} + \sigma_\mathrm{syst}^{2}}},
\end{equation}
where $f_\mathrm{obs}$ and $f_\mathrm{cal}$ are the points which come from the observation and model calculation; $\sigma_\mathrm{stat}$ and $\sigma_\mathrm{syst}$ are the statistical and systematic standard deviations of the observed points. This quantity could clearly show us the deviations between the best-fit result and observed values at each point based on its uncertainty.

The best-fit values, statistical mean values and standard deviations, and the 90\% confidence intervals for the parameters in three schemes are shown in Appendix Table \ref{tab:params_fitting}.

The fitting 1D probability and 2D credible regions (covariances) of posterior PDFs on the parameters of different schemes and groups are collected in Appendix \ref{app1}, \ref{app2}, and \ref{app3}. The data of the posterior samples of the parameter for three schemes is available on Zenodo under an open-source Creative Commons Attribution license: \dataset[doi:10.5281/zenodo.6435163]{https://doi.org/10.5281/zenodo.6435163}. 

\clearpage

\clearpage
\section{Covariances of Parameters in Scheme I}
\label{app1}

\begin{figure}[htp]
\centering
\includegraphics[width=0.5\textwidth]{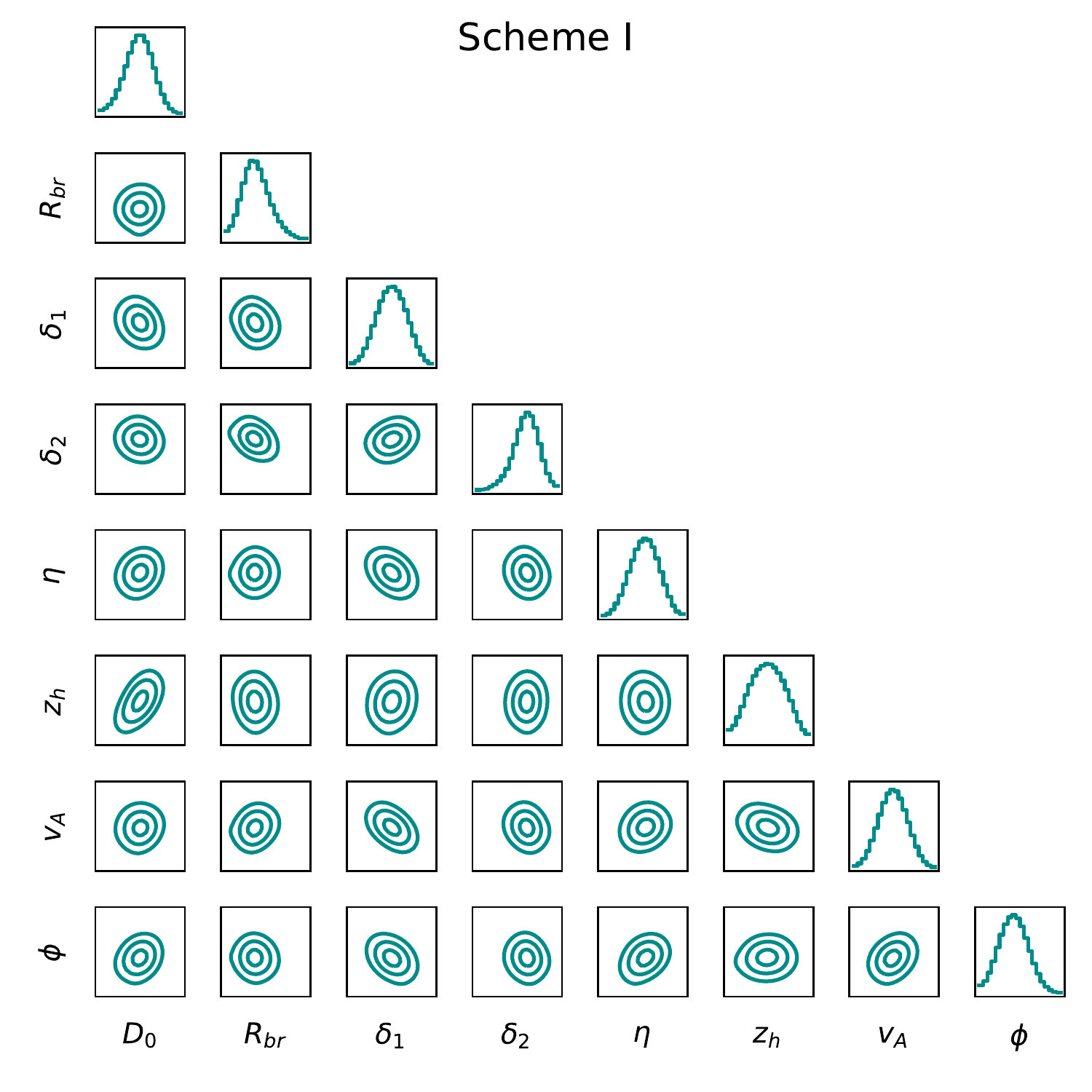}
\caption{Fitting 1D probability and 2D credible regions (covariances) of posterior PDFs on the propagation parameters in Scheme I. The contours present the $\sigma$, $2\sigma$ and $3\sigma$ CL.}
\label{fig:prop_1}
\end{figure}

\begin{figure}[htp]
\centering
\includegraphics[width=0.99\textwidth]{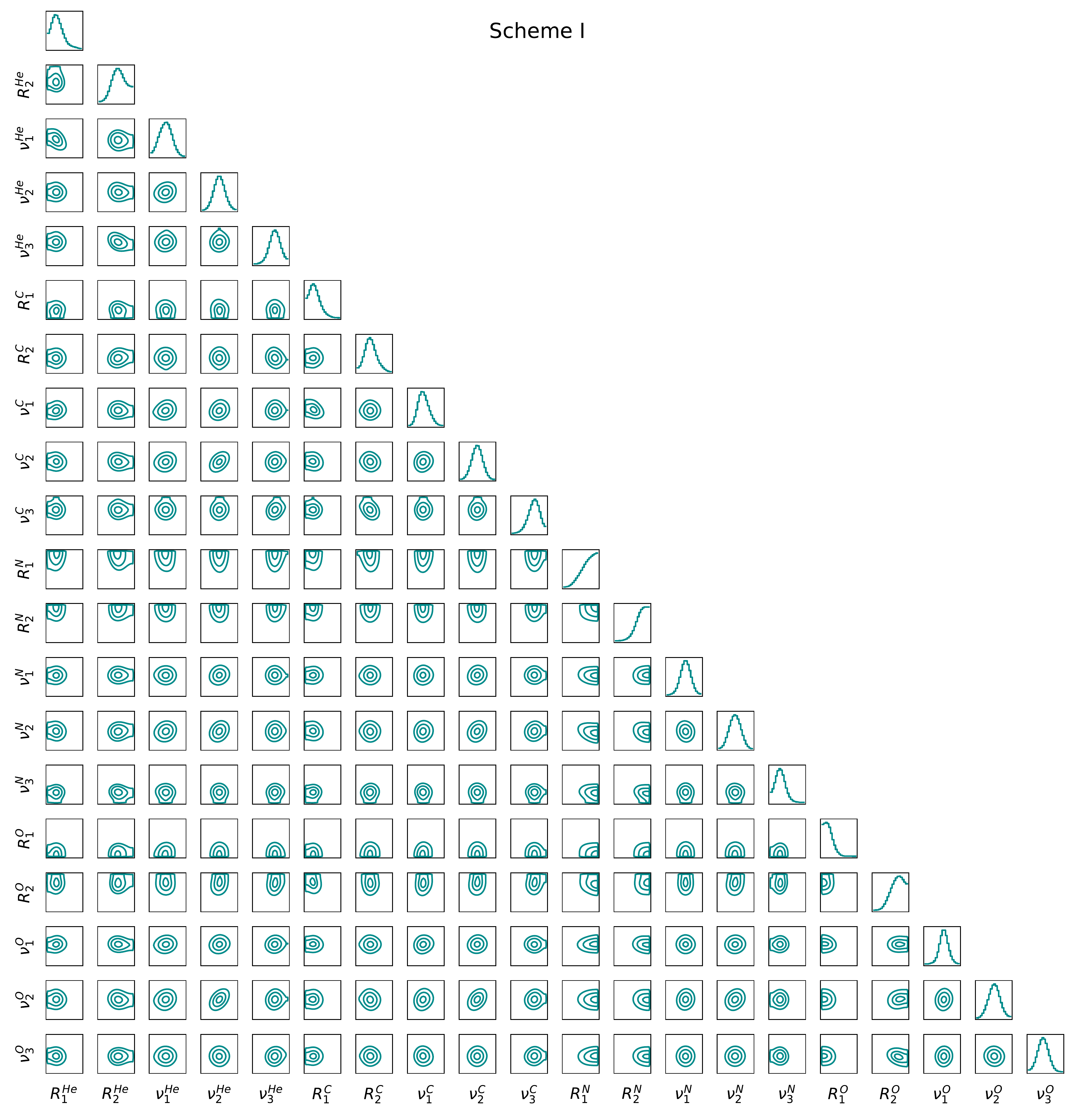}
\caption{Fitting 1D probability and 2D credible regions (covariances) of posterior PDFs on the He, C, N and O injection parameters in Scheme I. The contours present the $\sigma$, $2\sigma$ and $3\sigma$ CL.}
\label{fig:HeCNO_1}
\end{figure}

\begin{figure}[htp]
\centering
\includegraphics[width=0.9\textwidth]{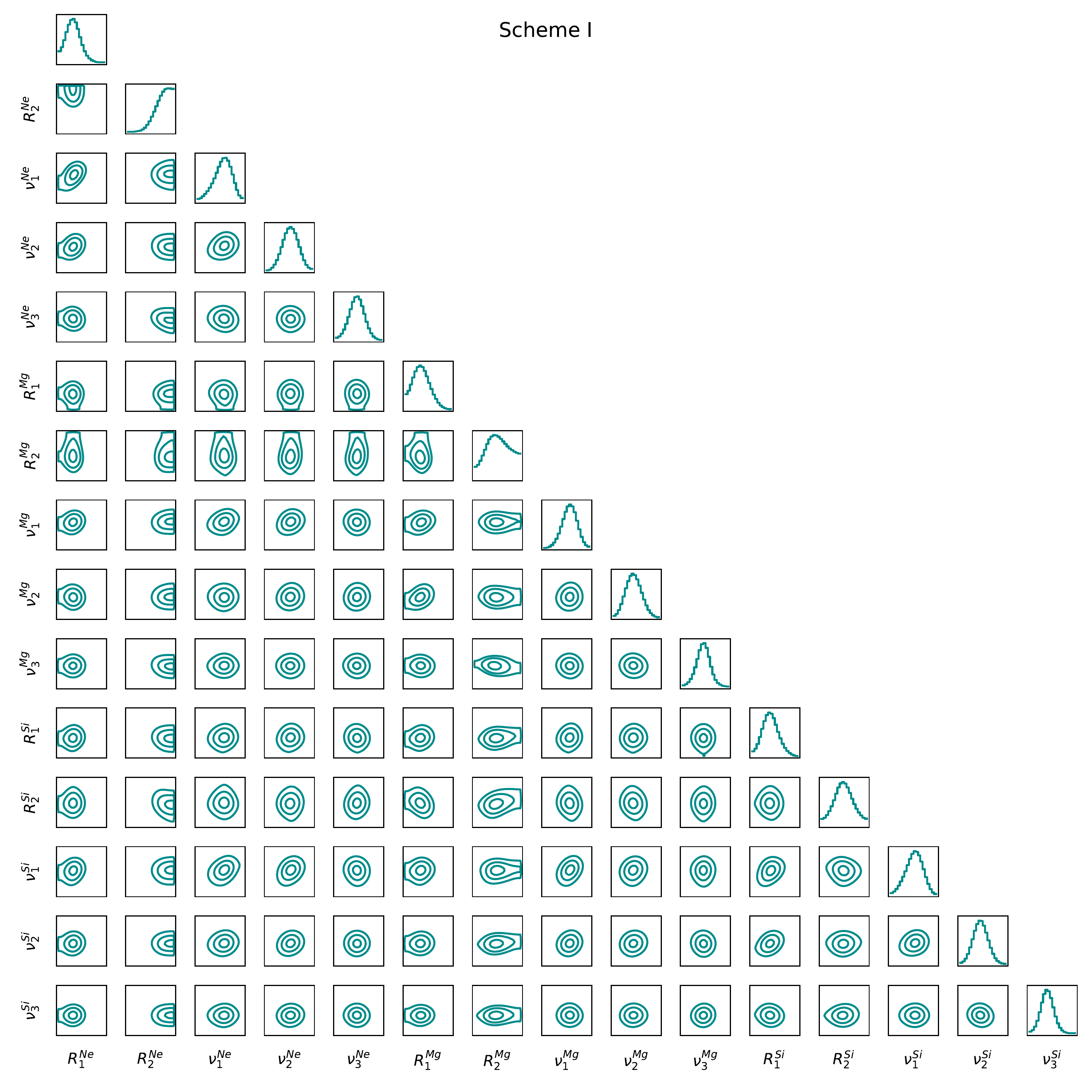}
\caption{Fitting 1D probability and 2D credible regions (covariances) of posterior PDFs on the Ne, Mg, and Si injection parameters in Scheme I. The contours present the $\sigma$, $2\sigma$ and $3\sigma$ CL.}
\label{fig:NeMgSi_1}
\end{figure}

\begin{figure}[htp]
\centering
\includegraphics[width=0.5\textwidth]{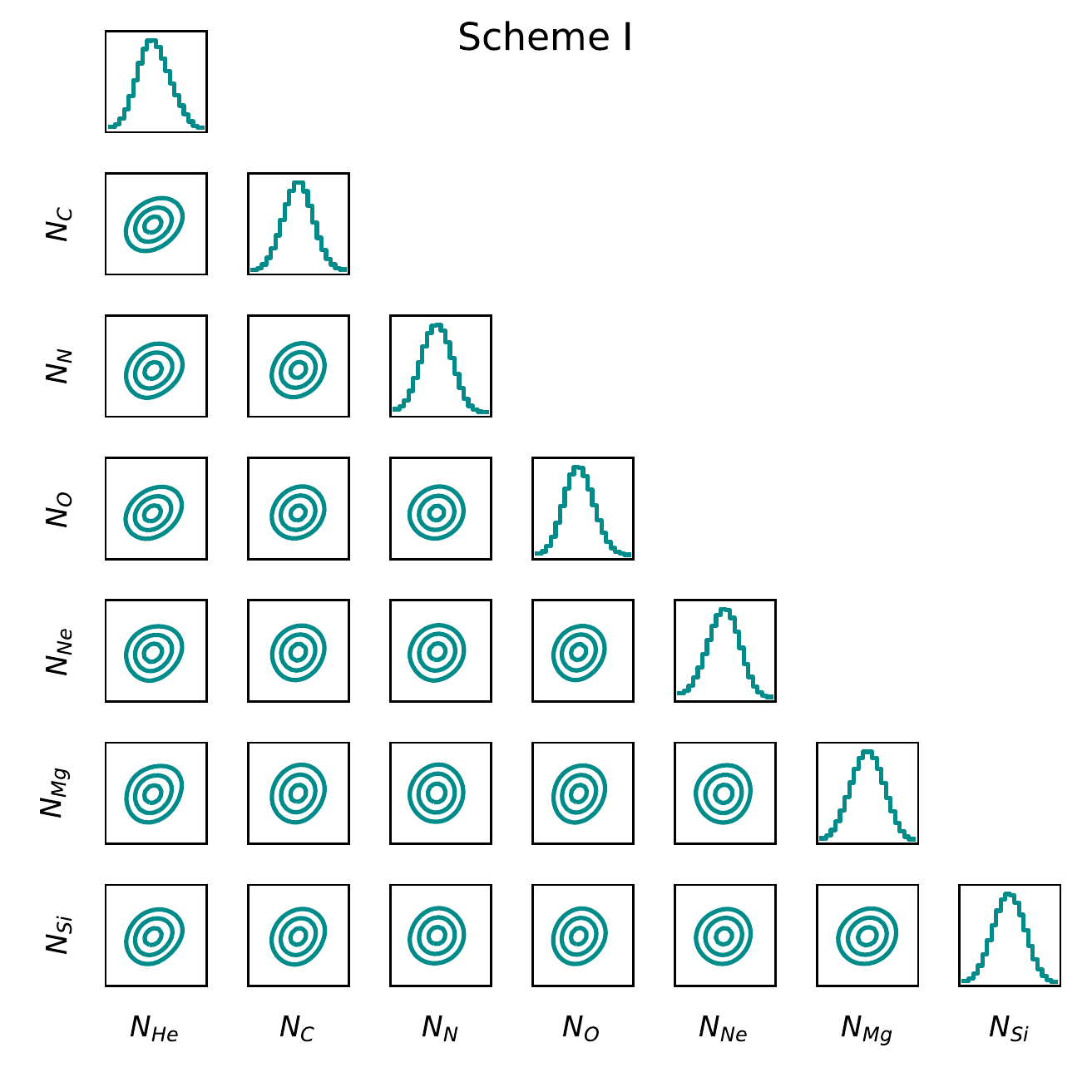}
\caption{Fitting 1D probability and 2D credible regions (covariances) of posterior PDFs on the primary source injection normalization parameters in Scheme I. The contours present the $\sigma$, $2\sigma$ and $3\sigma$ CL.}
\label{fig:norm_1}
\end{figure}

\clearpage
\section{Covariances of Parameters in Scheme II}
\label{app2}

\begin{figure}[htp]
\centering
\includegraphics[width=0.5\textwidth]{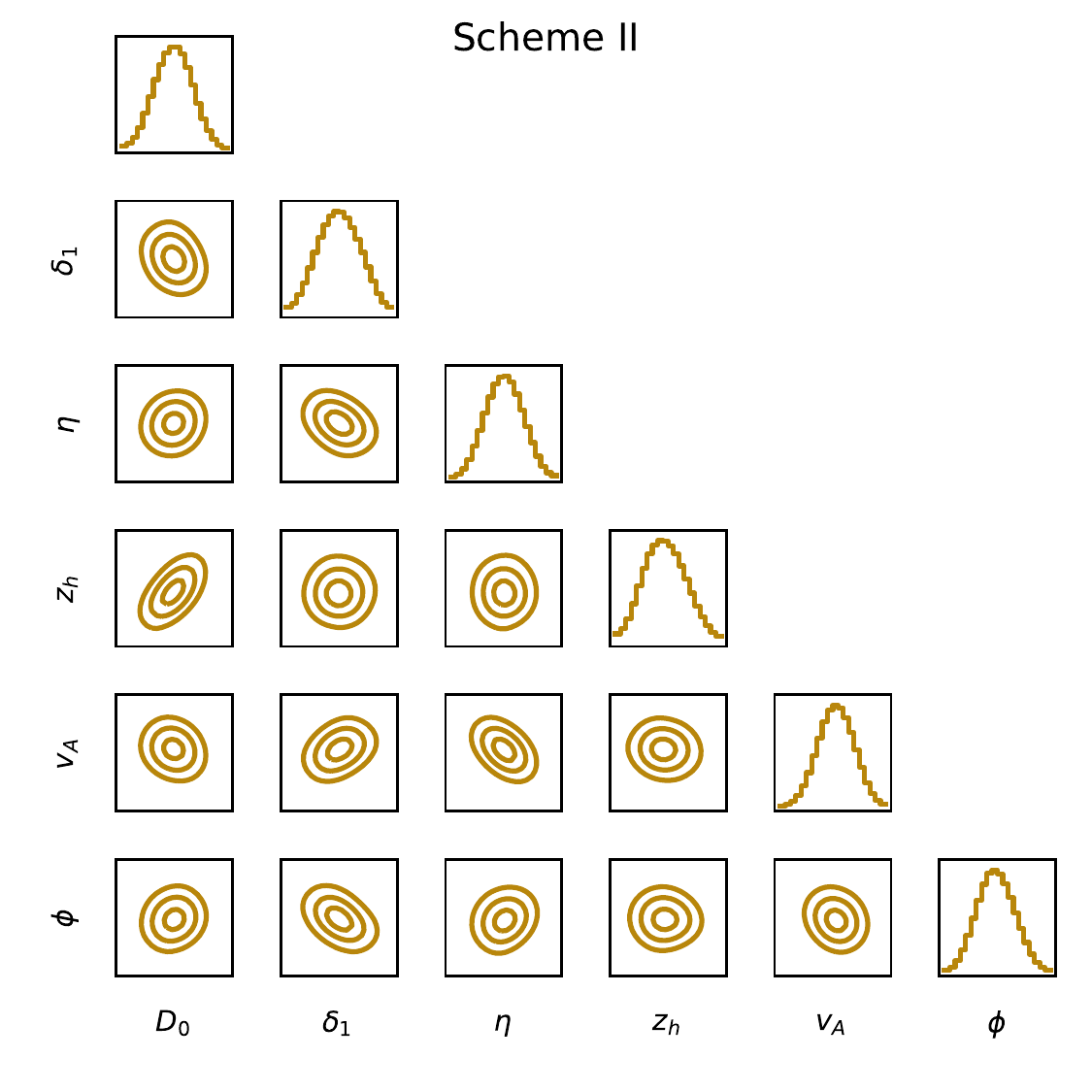}
\caption{Same as Fig. \ref{fig:prop_1} but for Scheme II.}
\label{fig:prop_2}
\end{figure}

\begin{figure}[htp]
\centering
\includegraphics[width=0.99\textwidth]{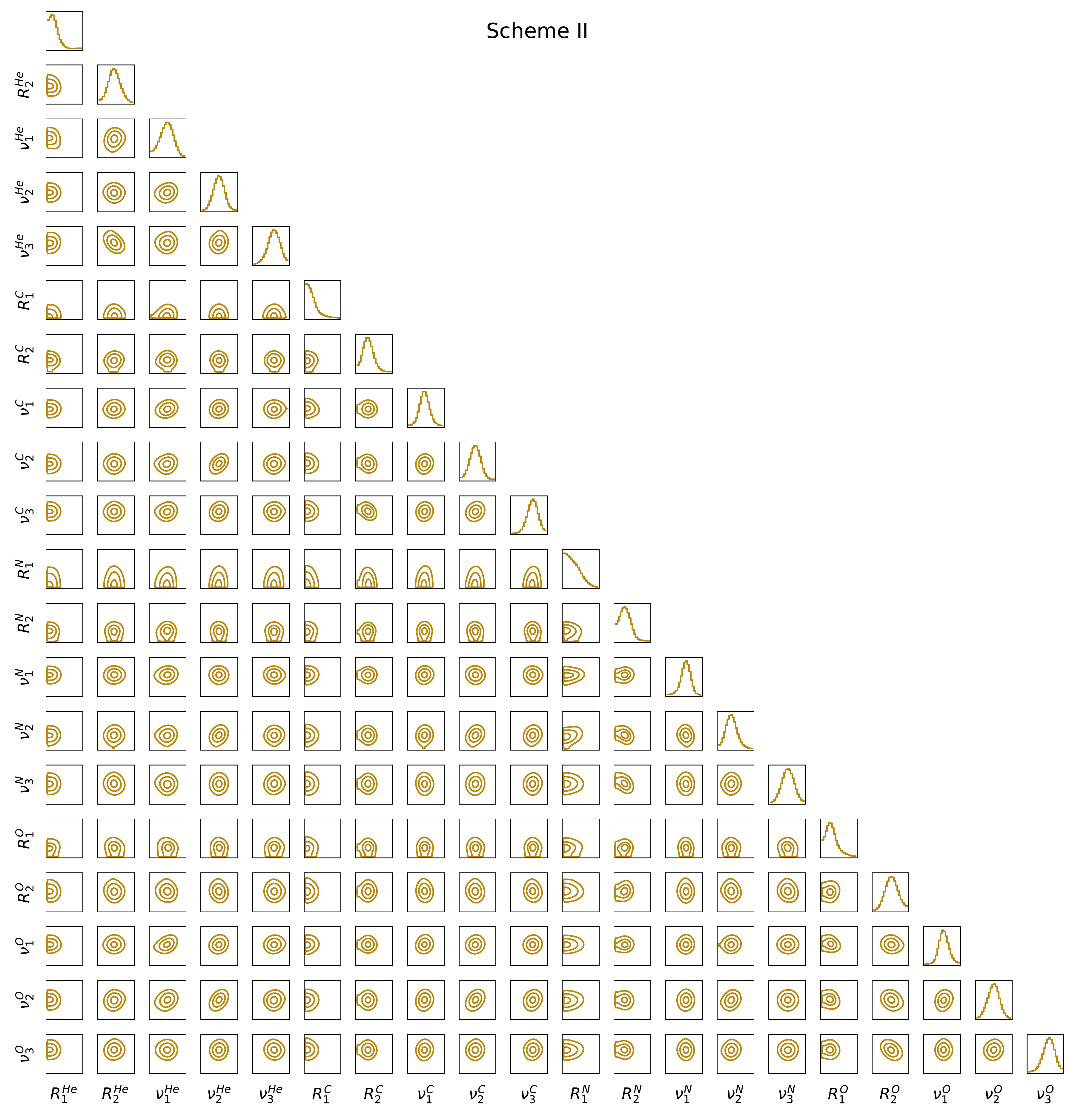}
\caption{Same as Fig. \ref{fig:HeCNO_1} but for Scheme II.}
\label{fig:HeCNO_2}
\end{figure}

\begin{figure}[htp]
\centering
\includegraphics[width=0.9\textwidth]{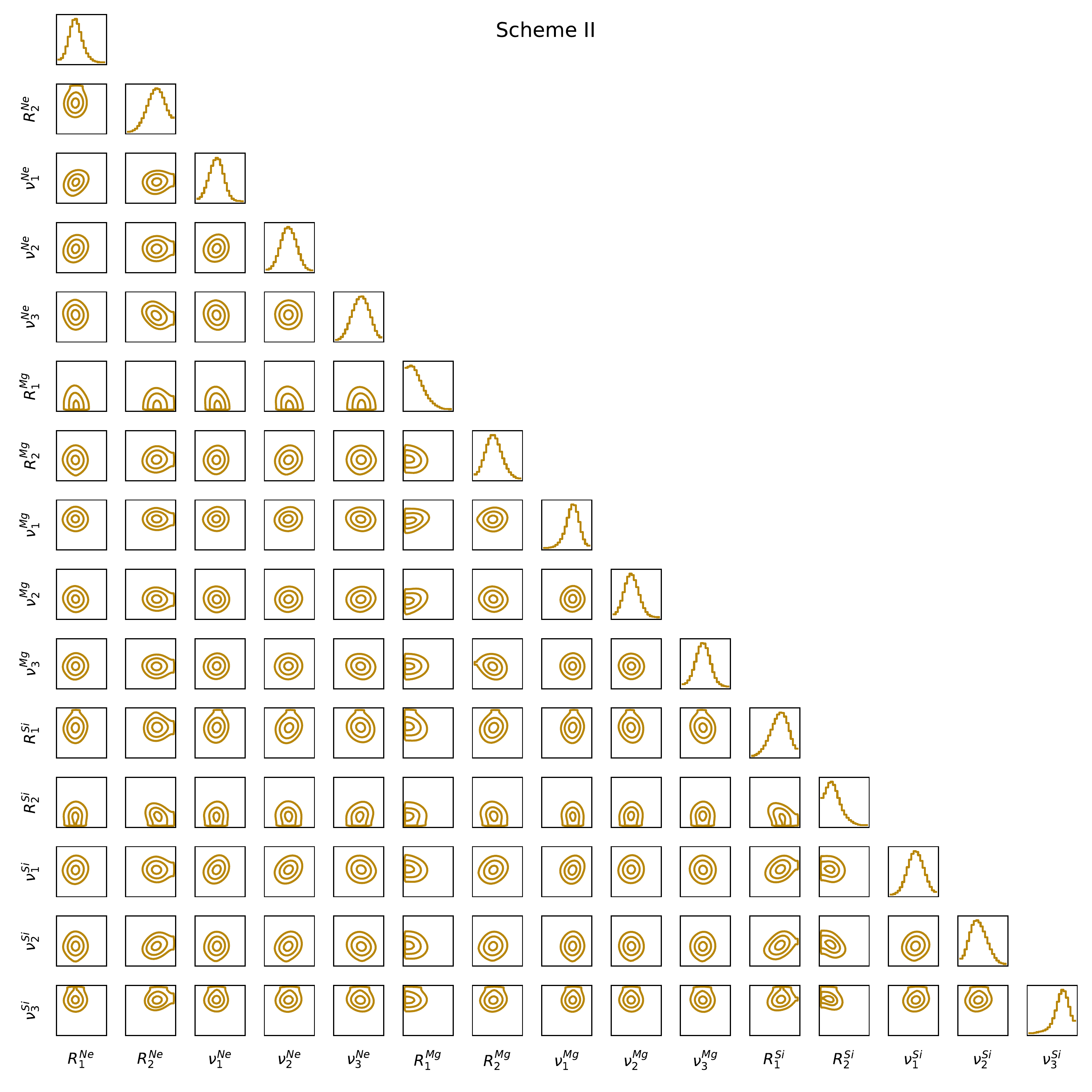}
\caption{Same as Fig. \ref{fig:NeMgSi_1} but for Scheme II.}
\label{fig:NeMgSi_2}
\end{figure}

\begin{figure}[htp]
\centering
\includegraphics[width=0.5\textwidth]{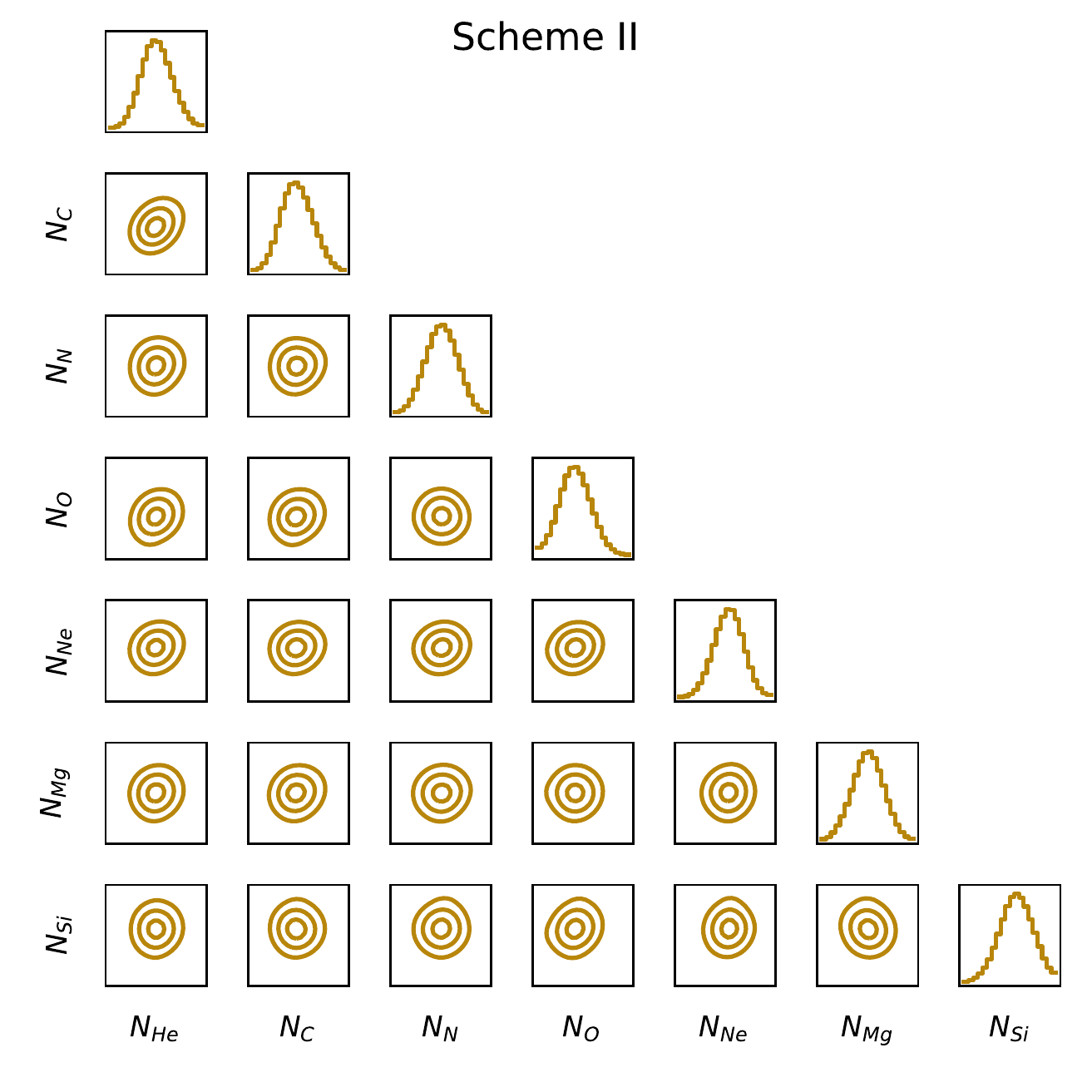}
\caption{Same as Fig. \ref{fig:norm_1} but for Scheme II.}
\label{fig:norm_2}
\end{figure}

\clearpage
\section{Covariances of Parameters in Scheme III}
\label{app3}

\begin{figure}[htp]
\centering
\includegraphics[width=0.5\textwidth]{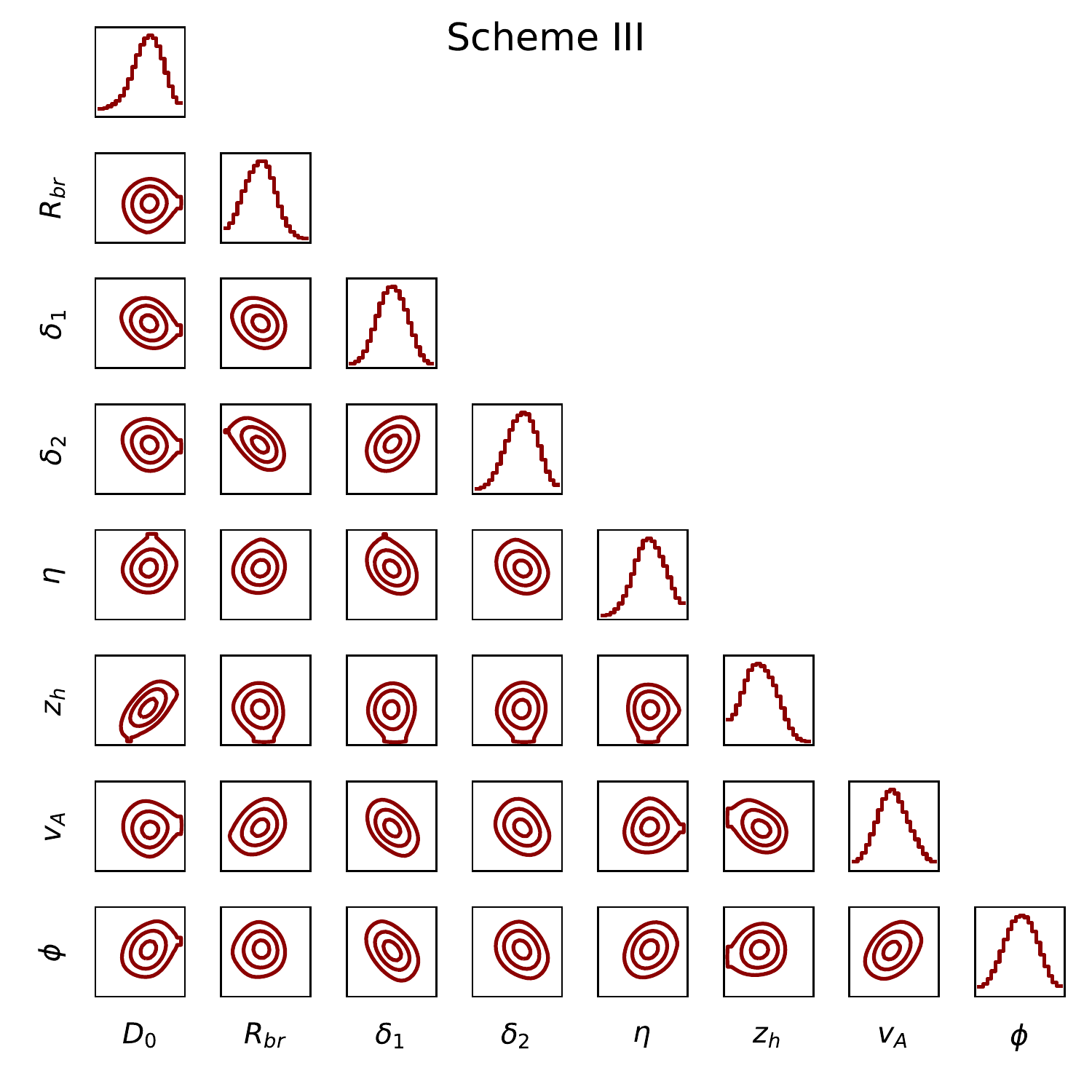}
\caption{Same as Fig. \ref{fig:prop_1} but for Scheme III.}
\label{fig:prop_3}
\end{figure}

\begin{figure}[htp]
\centering
\includegraphics[width=0.9\textwidth]{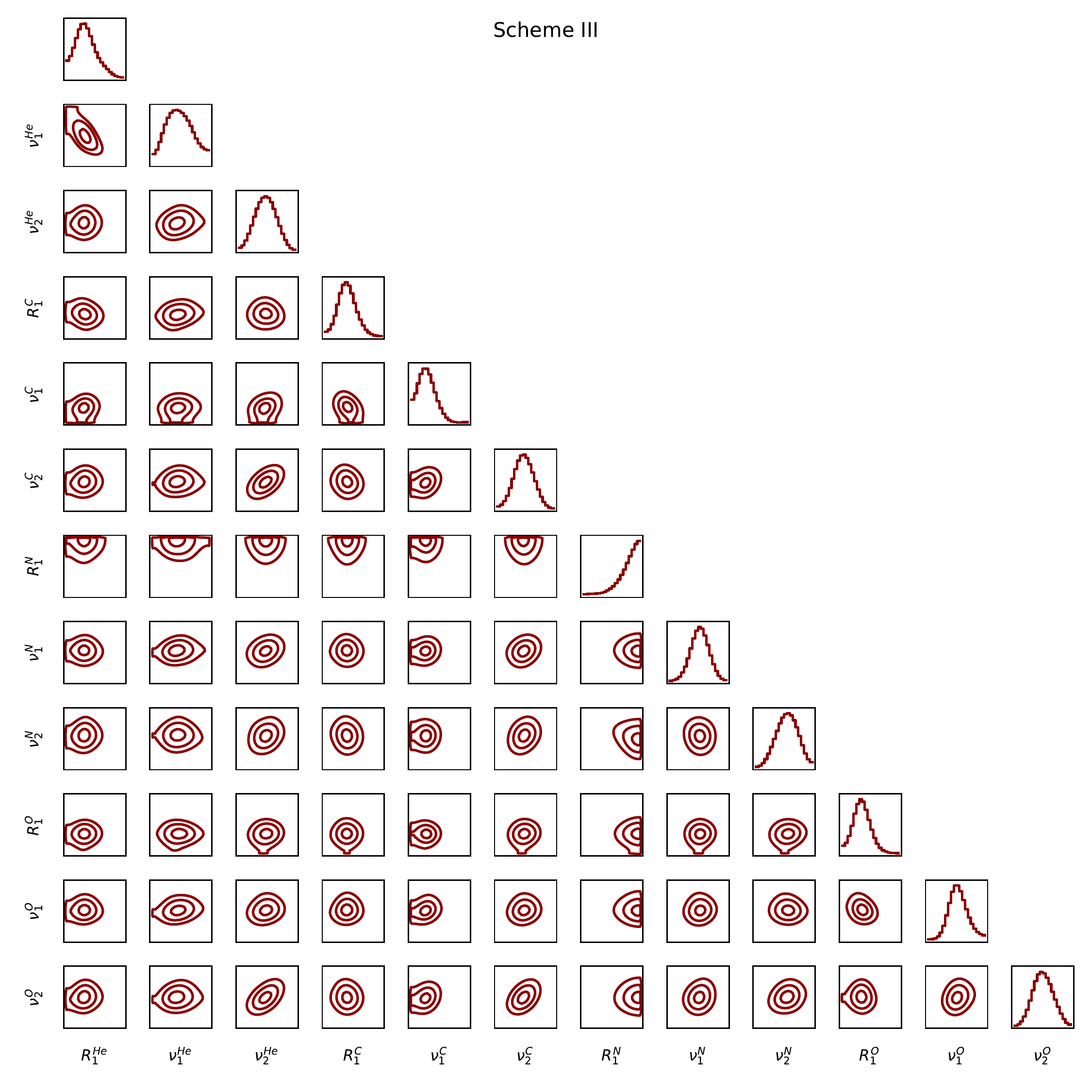}
\caption{Same as Fig. \ref{fig:HeCNO_1} but for Scheme III.}
\label{fig:HeCNO_3}
\end{figure}

\begin{figure}[htp]
\centering
\includegraphics[width=0.5\textwidth]{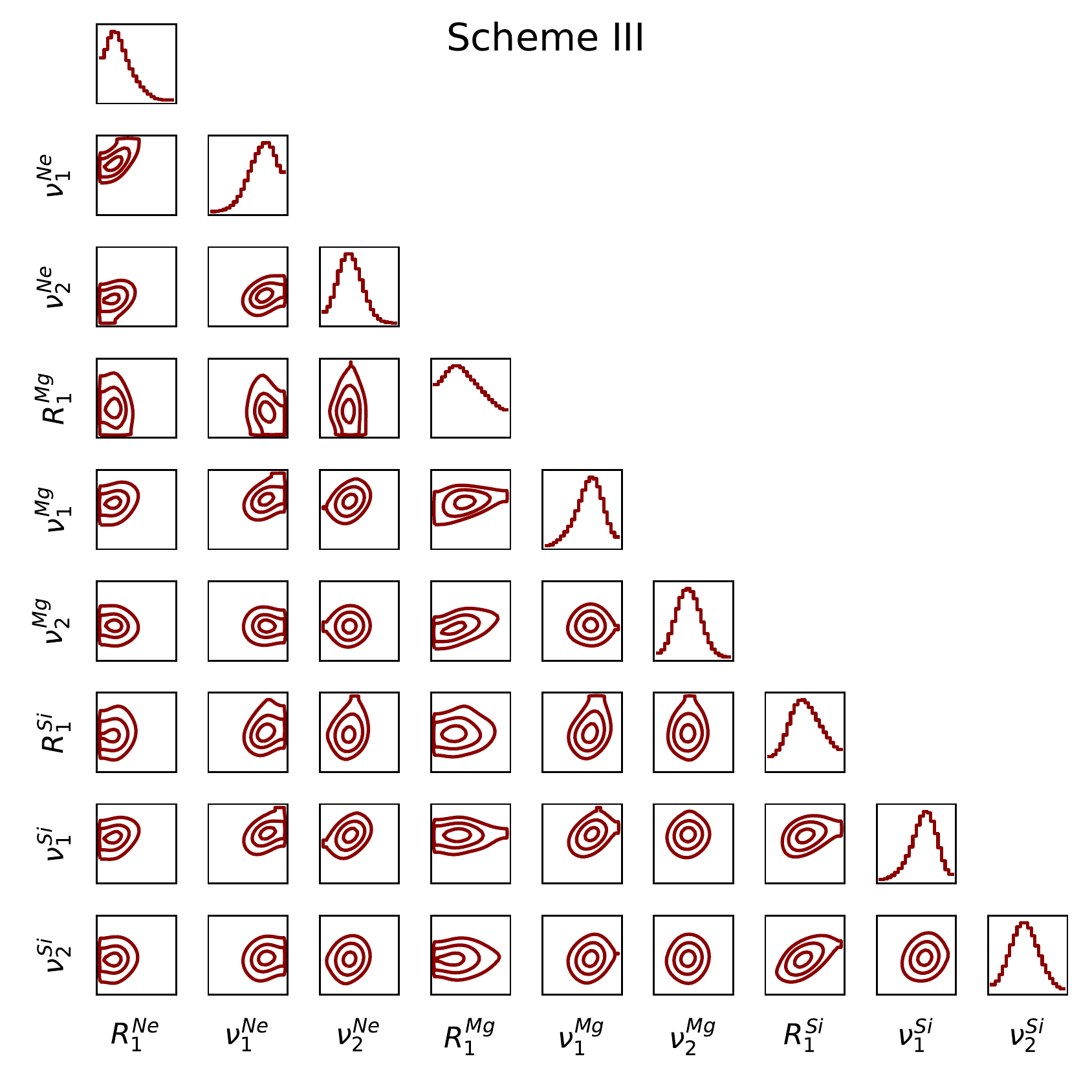}
\caption{Same as Fig. \ref{fig:NeMgSi_1} but for Scheme III.}
\label{fig:NeMgSi_3}
\end{figure}

\begin{figure}[htp]
\centering
\includegraphics[width=0.5\textwidth]{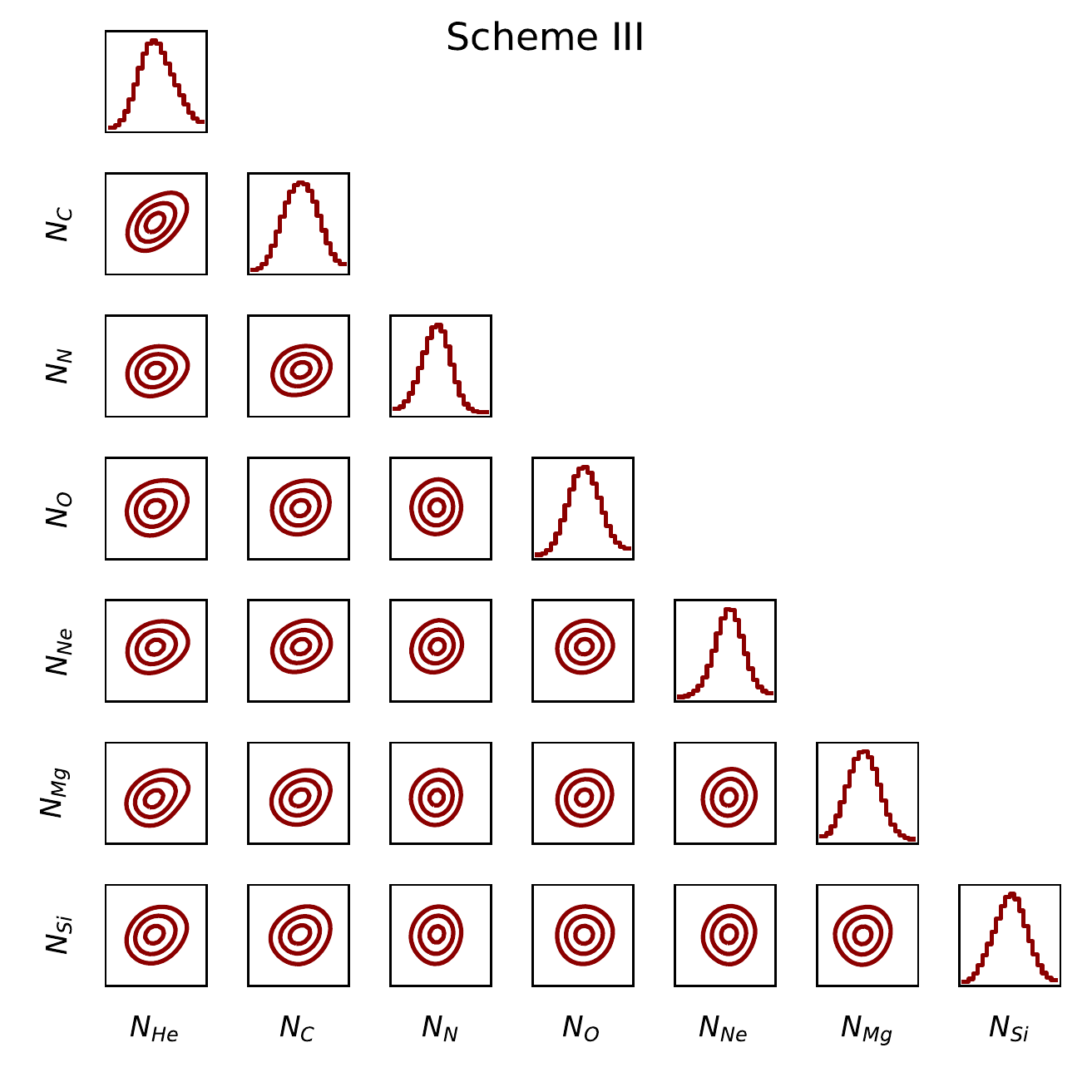}
\caption{Same as Fig. \ref{fig:norm_1} but for Scheme III.}
\label{fig:norm_3}
\end{figure}

\clearpage


\end{CJK*}
\end{document}